\documentclass[10pt,a4paper]{article}
\usepackage{graphpap,epsfig,amssymb,graphicx,textcomp,xcolor, amsmath}
\usepackage{multirow}
\usepackage{multicol}
\usepackage{hyperref}

\begin{document}
\begin{center}

{\LARGE {\bf Multiple crossover in the decay of metastable volume fraction of a Blume-Capel ferromagnetic needle}}

\vskip 0.5cm

{\large {\it Ishita Tikader$^1$ and Muktish Acharyya$^{2,*}$}} \\

{\it Department of Physics, Presidency University,}\\
{\it 86/1 College Street, Kolkata-700073, INDIA}\\
{$^1$E-mail:ishita.rs@presiuniv.ac.in}\\
{$^2$E-mail:muktish.physics@presiuniv.ac.in}

\end{center}
\vskip 1cm
\noindent {\bf Abstract:} The transient behaviours of a Blume-Capel ferromagnetic needle have been studied extensively by Monte-Carlo simulation. The needle has an elongated length in one direction compared to its cross-section. In the context of transient behaviour, we have captured the decay of the metastable state and the magnetic relaxation behaviours in our study. The dependence of metastable behaviour with anisotropy (single site) has been studied. \emph{Interestingly, we have observed multiple (different values of $n$ in different time domains) crossover in the decay of metastable volume fraction (obeying Avrami's law $\beta \sim {\rm exp}(-Kt^n)$). We have identified the crossover time, and the values of $n$ are estimated precisely.} The mean (magnetic) reversal time has been studied as a function of anisotropy. It is observed to be almost independent of anisotropy for its negative value; however, it is found to decrease exponentially with positive anisotropy. The exponential relaxation behaviour (in the corresponding paramagnetic phase) is observed. The relaxation time was found to depend on the strength of anisotropy. 
\vskip 2cm


\vskip 1cm


\vskip 1cm

\noindent $^*$ Corresponding author
\newpage

\section{\bf {Introduction:}}
The relaxation phenomenon is a well-known dynamical feature observed in physical systems. In magnetic systems, relaxation has been extensively investigated \cite{rlx}, motivated by its practical relevance in memory effects and medical applications. More recently, a brief review on the relaxation of magnetisation in model spin systems has been presented \cite{ishita-rev}. The metastable magnetisation and its decay are another interesting example of transient behaviours in magnetic systems \cite{becker, stauffer, naskar-rev}. The surface-bulk competitive metastable behaviour has recently been studied \cite{surface-bulk} by Monte Carlo simulation. 

The decay of the metastable volume fraction is generally governed by Avrami's law \cite{K,JM, A1, A2, A3}. The metastable volume fraction decays as $\beta \sim {\rm exp}(-Kt^n)$, usually referred to as Avrami's law. The factor $K$ is constant, and the power of $t$ (time) is $d+1$ in the bulk material, where $d$ is the dimension of the system or the dimension of growth. This law is verified \cite{NA, NA2, NAVF} recently in the decay of metastable states in the Ising and Blume-Capel models by extensive Monte Carlo simulations. No deviation from Avrami's law has been detected in such bulk materials. The powered exponential decay has been observed with a unique value of the power $n$.

A series of studies can be cited here where the crossover in decay or growth of volume fraction has been found in different systems. The rate is observed to change in different time domains in the case of glass-forming liquids \cite{slade1} in alkane and alcohols \cite{slade2}. A slow crossover from super-diffusion to diffusion has recently been observed \cite{gopal} in an isotropic spin chain. The crossover from anomalous to normal diffusion is found \cite{daniel} in lipid bilayers. The two-tier crystallisation regime in hybrid and nanocomposite polymers has been explored \cite{jabbar} using molecular dynamics simulations. The crossover from glass-like to liquid-like molecular diffusion has been observed recently \cite{ranieri} in a supercritical fluid. 

\textcolor{blue}{The decay of the metastable state is an important field of research in magnetic models. It has been widely studied in Ising ferromagnet. However, the behaviour of metastability and the deacy of metastable states in Blume-Capel model have not been
explored in that rigour. In the Blume-Capel model the single site anisotropy plays very important role in equilibrium phases. 
The Spin-1 Ising Blume-Capel model\cite{blume,capel} and its generalization the Blume-Emery-Griffiths model (BEG)\cite{beg} are
useful for representing a variety of physical and chemical
systems. Historically, These models have been introduced to describe the behaviours of superfluidity and the
phase separation in $He^3-He^4$ mixtures\cite{beg}. To describe the properties
of many substances, variant of these models are used\cite{variant}. Different techniques are employed to investigate the
equilibrium behaviours of the Blume-Capel model. Briefly, effective field theory with differential operator technique\cite{eff1,eff2,eff3},
Monte-Carlo simulation (MC)\cite{mc}, transfer-matrix finite-size-scaling (TMFSS)\cite{tmfss}, series expansion method
\cite{se1,se2}
 are employed.} \textcolor{red}{Some very recent important works on Blume-Capel model would be worthmentioning here. Several critical characteristics have been explored\cite{macedo} in anisotropic Blume-Capel and Baxter-Wu model from the Monte Carlo study of the distributions of energy and magnetisation.  The critical properties of small scale two dimensional Blume-Capel model have been studied\cite{moueddene} by Monte Carlo methods which are comparable with that obtained from the analysis of zeros of the
 partition function. Monte Carlo simulation (with Wang-Landau algorithm)
 has been employed\cite{mataragkas1} to explore the phase diagram
 (in temperature - crystal field plane) of two dimensional Blume-Capel model on triangular lattice. The infinite strip of Blume-Capel
 model on triangular lattice has been studied\cite{mataragkas2} by transfer matrix method to locate the tricritical point accurately. This
 precise estimate agrees well with that obtained from the conformal field theory predictions for the tricritical Ising universality class.}

\textcolor{blue}{How does the metastable state decay in a Blume-Capel ferromagnetic needle? A question naturally arises
why should we consider a needle-like object ? How does the behaviour of such quasi-one dimensional object differ from that of a bulk sample ? In this context, let us cite a series of references which shows that the quasi-one dimensional systems behave unusually different from that of bulk object. The hard
rodlike objects are found \cite{gurin} to form long chains (in low density) revealing the first order phase transition. The
quasi-one dimesional (or chain like) structure of antiferromagnetic $RbCoCl_3$ at low temperature showed\cite{cottam} the role
of spin-phonon interaction in providing a temperature-dependent contribution for the frequencies of the $E_{1g}$ and $E_{2g}$ symmetric phonons that occur with frequencies comparable to those of the spin wave excitations (magnons) in this bulk compound.
The enhanced and unusual insulating state at high pressure has been found\cite{sereika} from the enlarged 
distortion of the $OsO_6$. The unusual magnetic anisotropy has been observed\cite{he} in quasi-one dimensional $BaCo_2V_2O_8$ 
chain in the room temperature paramagnetic phase. The specific heat and susceptibility measurements suggested\cite{matsushita}
that $CeCo_2Ga_8$ is a rare example of quasi-one dimensional Kondo lattice. 
The unusual pressure induced superconductivity (above 20 GPa) has been found\cite{zhao} in quasi-one dimensional PdTeI materials.
The  results of dissipative solitons are
observed\cite{descalzi} in quasi-one dimension.}

\textcolor{blue}{Keeping the facts, of the unusual behaviours of the quasi-one dimensional system in mind, we have studied the transient behaviours such as decay of metastable states, magnetisation reversal and the relaxation behaviours of a Blume-Capel ferromagnetic needle (the size of its height is much larger than the cross-sectional dimension) by extensive Monte-Carlo simulation \cite{binder}.} We indeed found multiple crossovers in the decay of metastable volume fraction. This result differs from the behaviours of the bulk system in the same model. In this paper, we report our results in the following arrangements: The model is introduced in the next section (Section 2). Section 3 is dedicated to describing the Monte-Carlo simulation scheme \cite{binder,newman}. The results (with figures and tables) are provided in Section 4. The paper ends with the concluding remarks in Section 5.

\section{\bf The Model:}
The spin-1 Blume-Capel model is described by the following Hamiltonian.
\begin{equation}
\label{hamiltonian}
H=-J\sum_{<i,j>}S_i^z S_j^z + D\sum_{i}{(S_i^z)}^2 - h_{ext} \sum_{i}S_i^z
\end{equation}

Where $S_i^z$ is the $z$ component of the spin variable at the $i^{th}$ site. Since $S=1$, the z-component of the spin variable, i.e., $S_i^z$ can take any one of the values, +1, 0 and -1. The first term of the Hamiltonian describes the exchange interaction between nearest neighbour (NN) spins (symbolised by $<i,j>$). Here, $J>0$ defines the uniform ferromagnetic coupling. The second term signifies the effect of single-site uniaxial anisotropy (or magneto-crystalline anisotropy) that arises due to the crystal field. $D$, the anisotropy parameter (also termed as crystal field coupling) is measured in unit of $J$. We have considered uniform strength of anisotropy throughout our study. The last term represents the Zeeman energy that corresponds to the interaction with a uniform external magnetic field $h_{ext}$ (measured in unit of $J$).

In this study, we have considered the Blume-Capel model on a needle-like system whose effective geometry is quasi-one-dimensional. The needle-like system has a highly extended longitudinal dimension along the $z$-axis in comparison to the significantly smaller cross-sectional dimensions along the $x$ and $y$ - axes. The size of the system is specified by $L_x = 4$, $L_y = 4$ and $L_z = 200$ ($L_z \gg L_x, L_y$); having total lattice sites $N=L_x \times L_y \times L_z=3200$. The system is exposed to open boundary conditions in all three directions.

\section{\bf{Simulation Procedure:}}
Our investigations are performed by employing the Monte Carlo simulation scheme using the Metropolis single spin-flip algorithm. Starting from the initial configuration, a lattice site is randomly chosen among $N$ options, say $i$. The present spin state at that site $i$ is duly noted, i.e. $S_i^z(initial)$. The updated value may be in any of the three possible states (+1, 0 and -1). The final trial state of $S_i^z$ has been selected randomly from any of these three values with equal probability using a uniformly distributed random number ($r_1$) between 0 and 1. The trial spin state is labelled as $S_i^z(final)$. Now, the proposed spin update from the initial state $S_i^z(initial)$ to the final state $S_i^z(final)$ will be accepted based on the transition probability determined by the Metropolis rule\cite{binder,newman}.
\begin{equation}
\label{prob}
       P(S_i^z(initial) \rightarrow  S_i^z(final)) = Min \Biggl[ 1, \exp \biggl(-\frac{\Delta H}{k_BT} \biggr) \Biggr]  
\end{equation}
where $\Delta H$ is the change in energy caused by the change in spin state of $S_i^z$, from $S_i^z(initial)$ to $S_i^z(final)$. $\Delta H$ (in unit of $J$) is determined by the BC Hamiltonian, mentioned in equation (\ref{hamiltonian}). $k_B$ is the Boltzmann constant, and $T$ is the temperature of the system measured in unit of $J/k_B$. For simplicity, $J =1$ and $k_B = 1$ are assumed throughout our simulation. Now generate a random number $r_2$ uniformly distributed in the range of $[0,1]$ and compare with the transition probability $P(S_i^z(initial) \rightarrow  S_i^z(final))$ (determined by the equation (\ref{prob})). The trial update will be accepted if the $P(S_i^z(initial) \rightarrow  S_i^z(final)) \ge r_2$ condition is satisfied. Otherwise, the initial state of $S_i^z$ remains unchanged. The $N=3200$ randomly chosen lattice sites are updated following the aforementioned protocols. These $N$ random updates constitute one Monte Carlo step per spin (MCSS), the unit of time in our simulation. \\
At each MCSS, magnetisation of the system is determined by following,
\begin{equation}
\label{m(t)}
   m(t)= \frac{1}{N}\sum_{i} S_i^z  
\end{equation}
where $N=3200$ is the total number of lattice sites.\\
The time averaged magnetisation, $M=\langle m \rangle$. The magnetic susceptibility is calculated using $\chi = \frac{N}{k_BT}(\langle m^2 \rangle - {\langle m \rangle}^2)$. Specific heat is usually defined as the temperature gradient of internal energy, formulated as $C_v=\frac{\partial{\langle E \rangle}}{\partial T}$. $C_v$ is measured by the numerical differentiation of $\langle E \rangle$ with respect to temperature (using the central difference method), where $E$ is the energy per site determined from the equation (\ref{hamiltonian}).
Here $\langle .. \rangle$ denotes the time average of thermodynamic quantities, approximately equal to the ensemble average in the ergodic limit. 

\section{\bf{Results:}}
Our primary focus is to investigate the metastable behaviour of a Blume-Capel ferromagnetic needle of size $L_x = 4$, $L_y = 4$ and $L_z = 200$, where the cross-sectional area ($L_x \times L_y$) of the needle is very small compared to its length ($L_z$). In order to explore metastability in this system, we have to secure a rough estimate of the finite-size pseudo-critical point $T_c^p (D)$ associated with the second-order (ferro–para) phase transition in the needle-shaped Blume-Capel ferromagnet. A well-defined phase boundary is crucial to determine whether the system is in the ordered or disordered phase.

\begin{figure}[h!tpb]
 \centering
  \includegraphics[angle=0, width=1.00\textwidth]{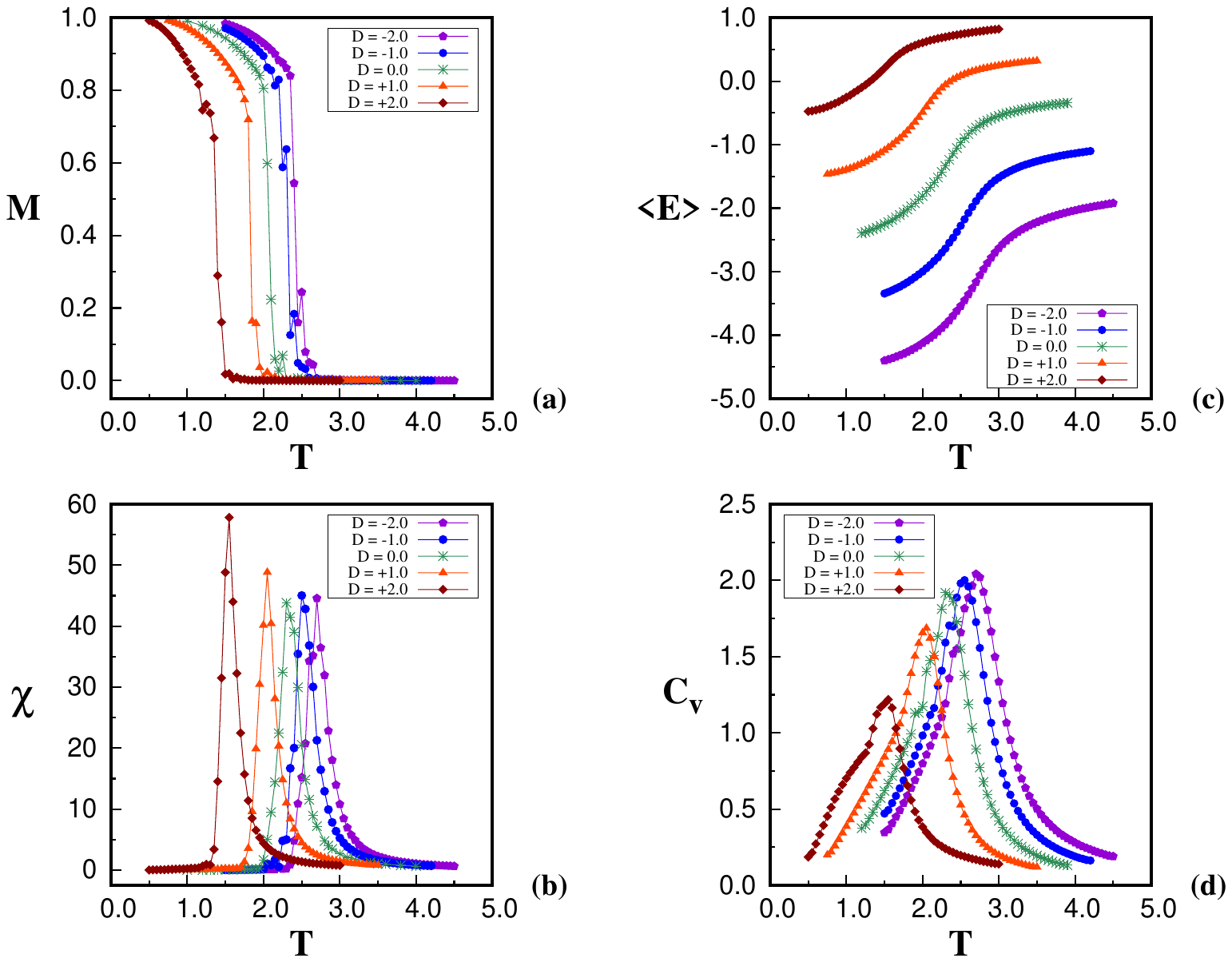}
  \caption{Temperature dependences of thermodynamic quantities; (a) Magnetisation, (b) Susceptibility, (c) averaged energy density, and (d) Specific heat in a Blume-Capel ferromagnetic needle of size $L_x = 4, ~ L_y=4,$ and $L_z=200$. Represented for different values of magnetic anisotropy $D = $-2.0, -1.0, 0.0, +1.0 and +2.0. Here, $h_{ext}=0$.}
  \label{fig:mag-eng}
\end{figure}

\begin{figure}[h!tpb]
 \centering
  \includegraphics[angle=0, width=1.00\textwidth]{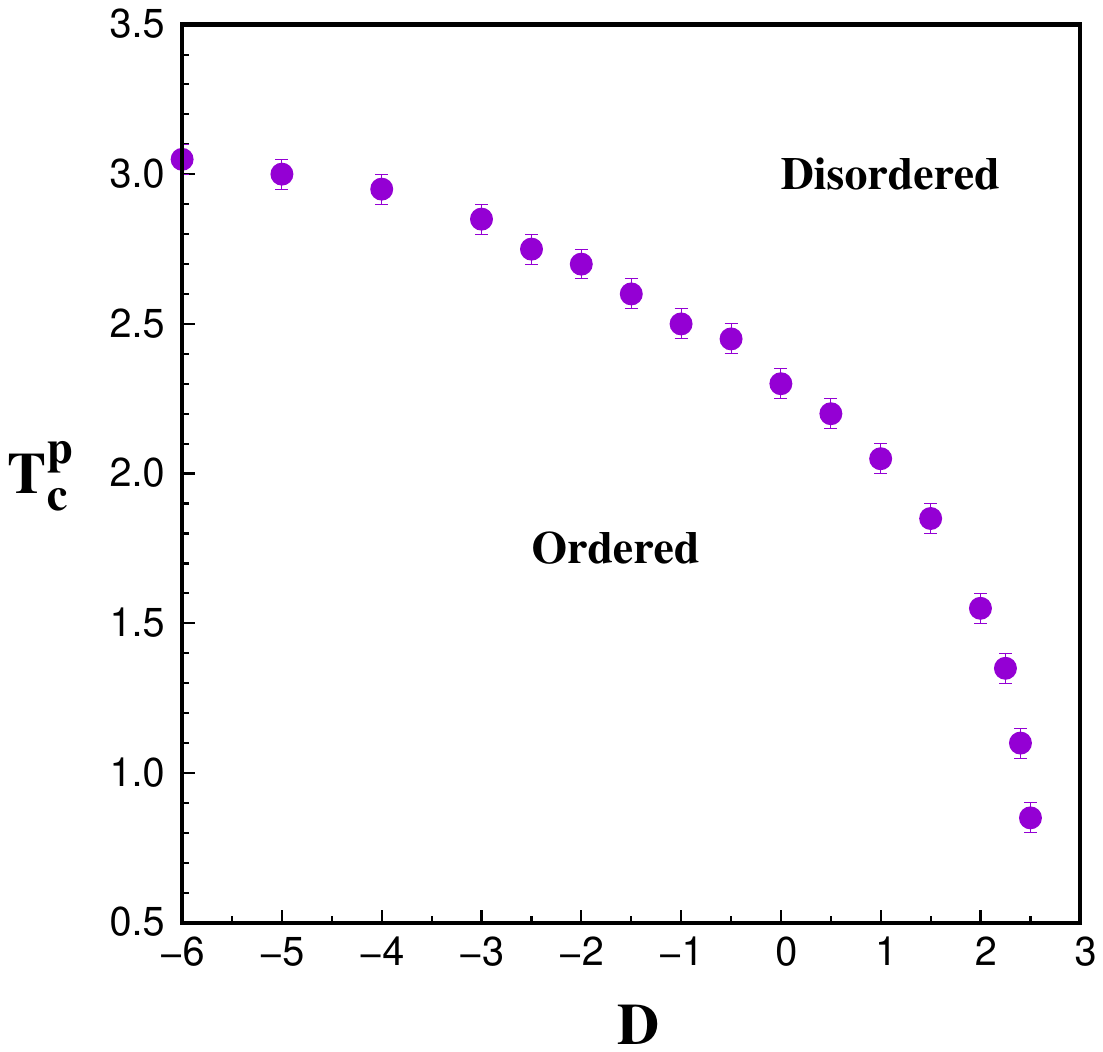}
  \caption{Phase diagram of a Blume-Capel ferromagnetic needle of size $L_x = 4, ~ L_y=4,$ 
  and $L_z=200$, on the $D-T_c^p$ plane. Here, $h_{ext}=0$.}
  \label{fig:phase-diagram}
\end{figure}

Starting from a high-temperature random spin configuration (disordered phase), the system is slowly cooled (for $h_{ext}=0$) in small temperature steps of $\delta T=0.05$. At each temperature step, the initial $10^4$ MCSS time steps are discarded to ensure thermal equilibrium, and the next $10^4$ MCSS are recorded to determine the time averages of the observables. Time-averaged quantities are further averaged over 1000 independent samples to minimise the fluctuations. The thermal variation of equilibrium thermodynamic quantities, i.e. time-averaged magnetisation $M$, susceptibility $\chi$, time-averaged energy density $\langle E \rangle$ and specific heat $C_v$ are critically studied for different strengths of uniform anisotropy in the range of $-6.0 \leq D \le +2.5$. and graphically illustrated in Fig. \ref{fig:mag-eng}. At the high-temperature regime, the random spin configuration of the system results in net zero magnetisation, called the disordered or paramagnetic phase. As we decrease the temperature of the system, the equilibrium magnetisation grows and achieves a non-zero value at the transition temperature, shown in Fig. \ref{fig:mag-eng}(a). The temperature dependence of magnetic susceptibility ($\chi$) reveals a sharp peak at a particular temperature, indicating the existence of a second-order phase transition, as depicted in Fig. \ref{fig:mag-eng}(b). The pseudo-critical temperature or transition temperature ($T_c^p$) is determined from a certain temperature that maximises susceptibility ($\chi$). Similarly, the specific heat ($C_v$) also exhibits a maximum at the transition temperature, as represented in Fig. \ref{fig:mag-eng}(d). Figure \ref{fig:mag-eng}(b) clearly shows that the susceptibility peaks shift towards lower temperatures as the anisotropy parameter $D$ increases. A similar shift in the specific heat peaks is observed, as shown in Fig. \ref{fig:mag-eng}(d). In order to elucidate the influence of single-site uniaxial anisotropy on the phase transition of the needle-shaped Blume-Capel ferromagnet, we obtain the phase diagram in the $D-T_c^p$ plane, as illustrated in Fig. \ref{fig:phase-diagram}.

\textcolor{blue}{It may be worth mentioning here that for a given interval of temperature ($\Delta T$), it would not be possible to detect any first order phase transition and the location of tricritical point. We believe that, like a cubic Blume-Capel ferromagnet, the needle would also show the first-order phase transition and the tricritical point.  To get the first order phase transition, a precise temperature variation  is required. Most importantly, the probability distribution of the order parameter has to be studied closely around the transition temperature. Moreover, as reported by the mean field \cite{mft1,mft2,mft3,mft4} and Monte Carlo studies\cite{FZTWJM, EZTF} that the first order line can be seen for precise variation of the anisotropy ($D$), which has not been explored in the present study. In particular, for the phase diagram, it may be an interesting study to explore the behaviours near $D=0$ and $T=0$. However, this is beyond the scope of present study.
A separate project may be undertaken to  investigate  the phase boundary of the Blume–Capel needle and thereby explore the tricritical behaviour.} 

 Figure \ref{fig:phase-diagram} clearly indicates that the pseudo-critical temperature $T_c^p$ decreases with increasing uniform anisotropy $D$. More specifically, for positive anisotropy ($D>0$), increasing the anisotropy strength ($|D|$) shifts the transition temperature $T_c^p$ to lower values. Conversely, for negative anisotropy ($D<0$), as the magnitude of anisotropy ($|D|$) becomes stronger, the transition point $T_c^p$ shifts towards the relatively higher temperature region. 

\begin{figure}[h!tpb]
 \centering
  \includegraphics[angle=0, width=1.00\textwidth]{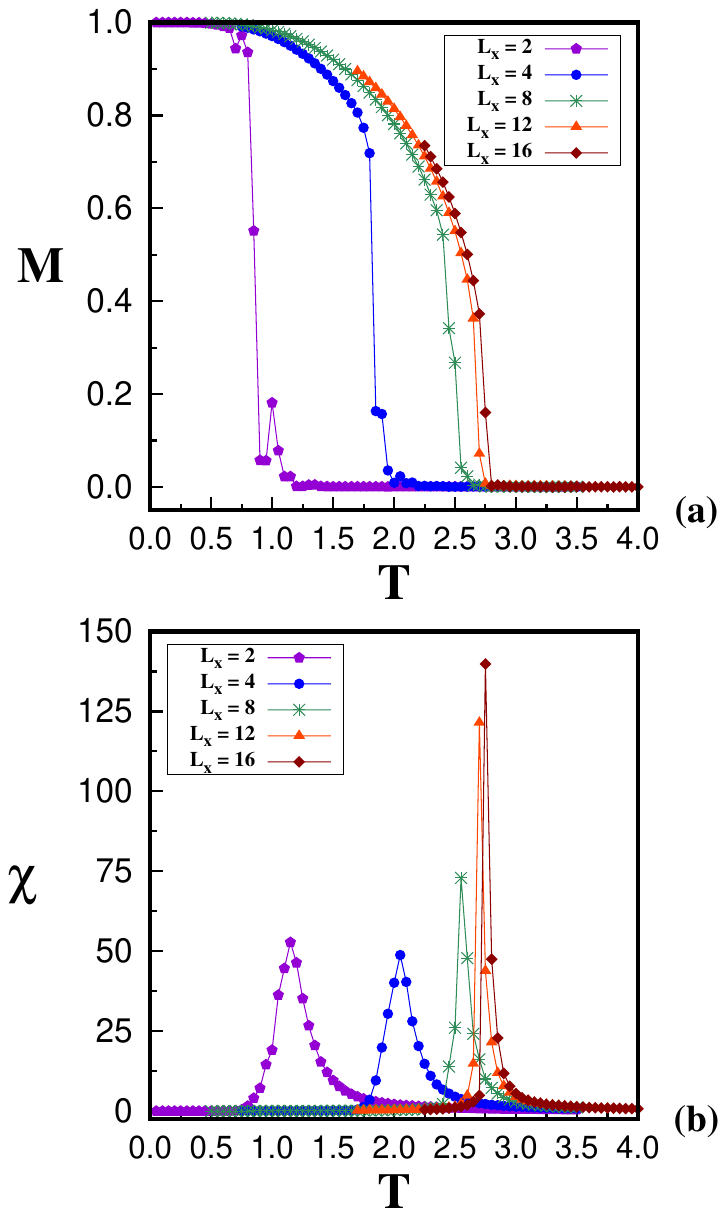}
  \caption{\textcolor{blue}{Temperature dependences of the (a) magnetisation and (b) susceptibility for different values of
  $L_x=L_y$ with fixed $L_z=200$, $D=+1$ and $h_{ext}=0$.}}
  \label{fig:diff-area}
\end{figure}

\begin{figure}[h!tpb]
 \centering
  \includegraphics[angle=0, width=1.00\textwidth]{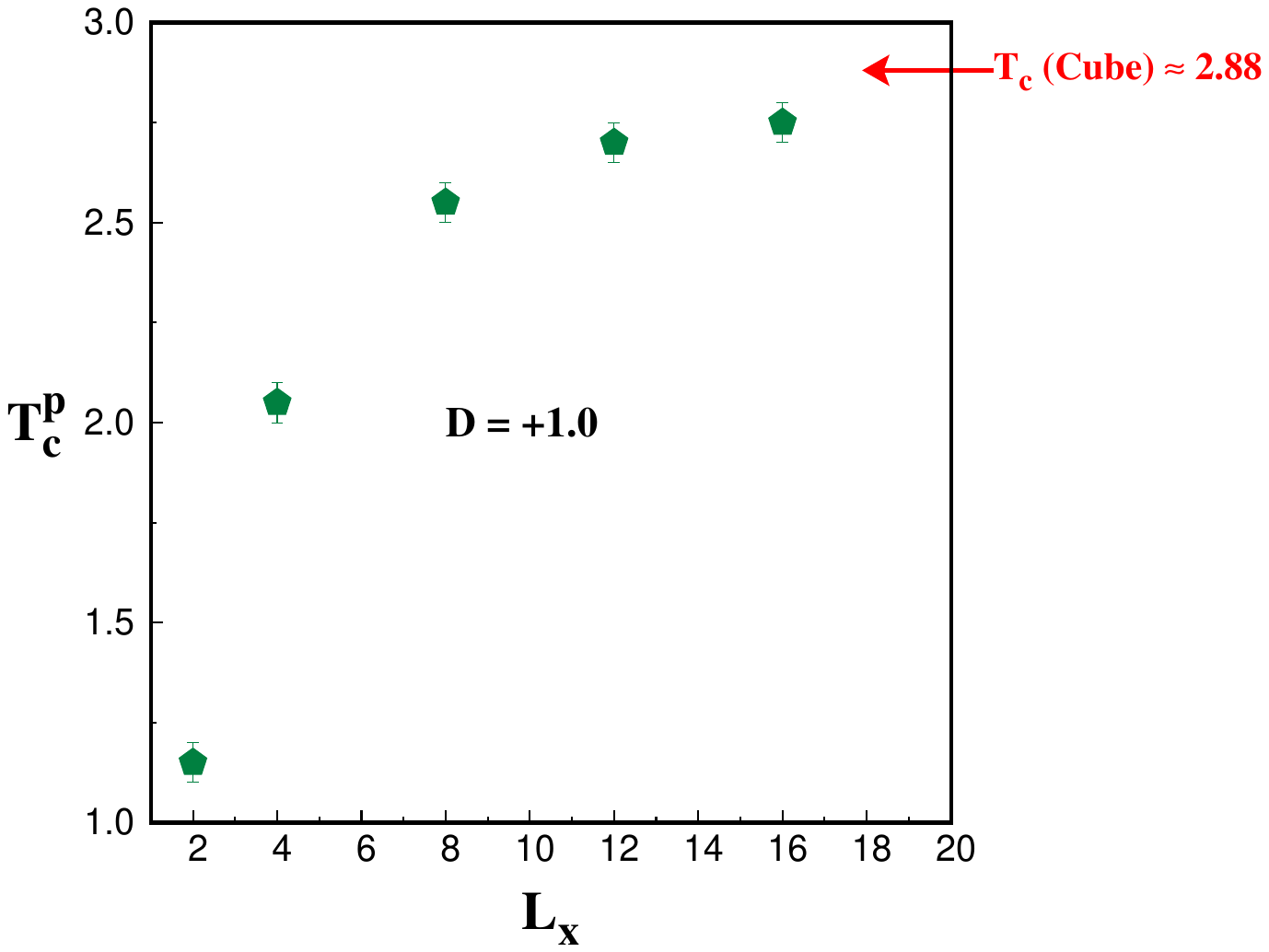}
  \caption{\textcolor{blue}{Variation of pseudo-critical temperature $T_c^p$ with  cross-sectional dimensions $L_x$ and $L_y$ (where $L_x = L_y$), for fixed values of $L_z=200$, $D=+1$ and $h_{ext}=0$. The pseudocritical temperature increases as the cross -sectional 
  area ($L_x \times L_y$) of the BC needle
  increases. The red arrow in right-top corner indicates the numerical estimates of pseudocritical temperature for Blume-Capel
  cube\cite{ozkan,kutlu}. }}
  \label{fig:Tc-area}
\end{figure}

\textcolor{blue}{A very important question may arise here. How does this pseudocritical temperature depend on the cross
sectional area ($L_x \times L_y$) of such BC needle? Even, it is not yet known what should be the strategy to study the
scaling behaviour of such system. In this regard, we have studied the equilibrium phase transition for different values
of $L_x=L_y$ (2, 4, 8, 12, 16) for fixed $L_z=200$. The results are shown in Figure \ref{fig:diff-area}. The results
shows that the critical temperature increases as the cross sectional area ($L_x \times L_y$) of the needle increases as shown in Figure \ref{fig:Tc-area}. 
In this way for $L_x=L_y=200$ the BC needle becomes a cube. The critical temperature has been found to reach the corresponding value for $D=+1$,  i.e. $T_c \approx 2.88$ \cite{ozkan,kutlu}. However, the systematic analysis of the scaling behaviour
with the cross sectional area would be definitely an interesting and important study, which requires huge computational efforts and may be
investigated in a separate project. Our objective is to study the transient behaviour only. However, we have investigated
the dependence of pseudocritical temperature for different values of cross sectional area keeping other parameters fixed.
}

\subsection{Metastable behaviour:}
The decay of a metastable state is a widely recognised and significant phenomenon in nature. A metastable state is defined as an intermediate energy state of a dynamical system, in which system remains for quite a long time before transitioning to the stable configuration.  Below critical temperature ($T<T_c$)  a ferromagnetic system can exist in one of two stable states: (i) with majority of spins either aligned up, giving positive magnetisation, or (ii) spins  predominantly down, yeilding negative magnetisation. When a weak magnetic field is applied opposite to the initial magnetisation, the system may become temporarily trapped in a metastable state.   The metastable state persists until thermal fluctuations and the applied field are sufficient to overcome the energy barrier of free energy and eventually decays towards the stable state with magnetisation aligned to the external field.

\begin{figure}[h!tpb]
 \centering
  \includegraphics[angle=0, width=1.00\textwidth]{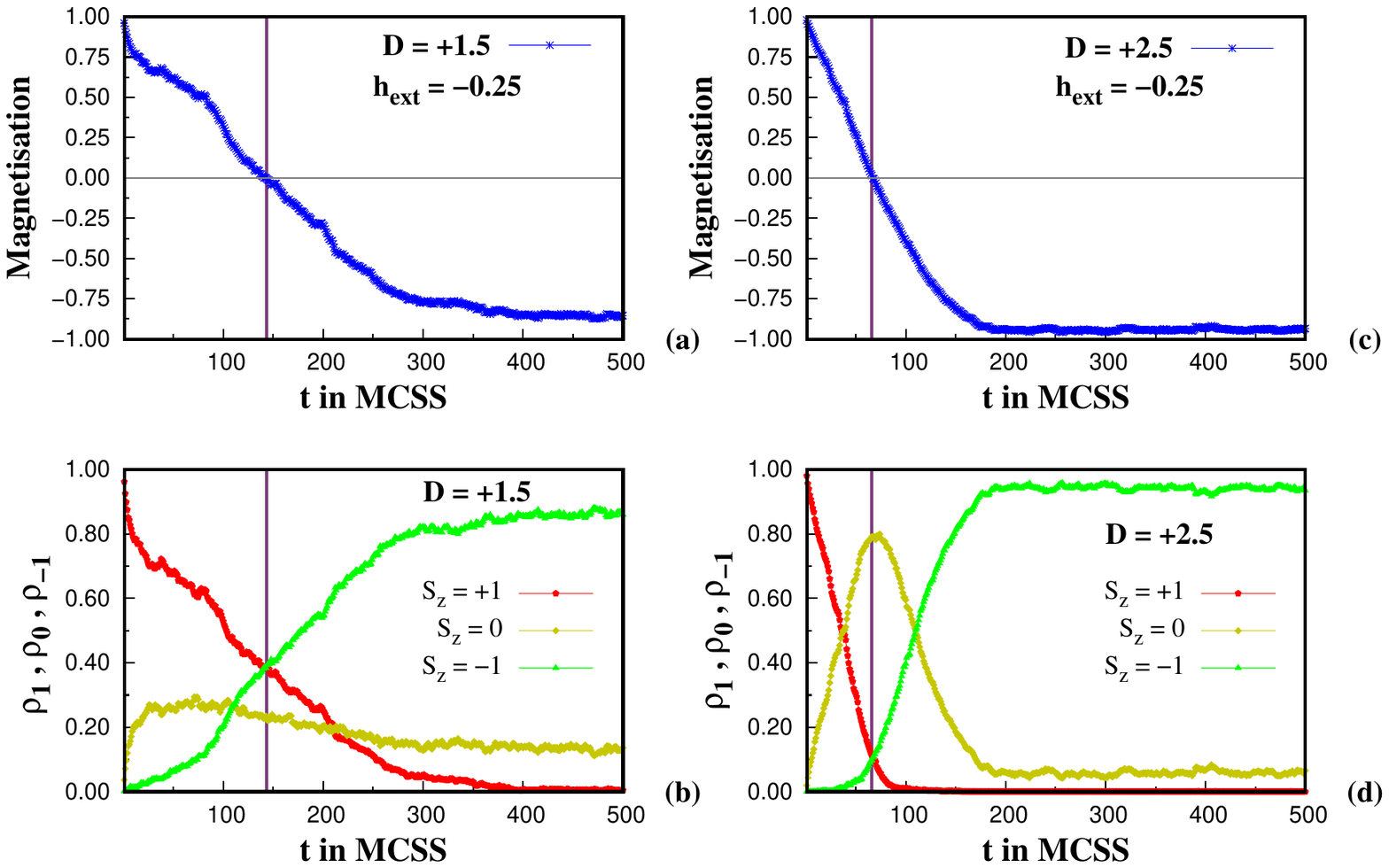}
  \caption{Top: Variation of magnetisation with time (in unit of MCSS) for two different values of positive anisotropy $D=$+1.50 and +2.50. The BC ferromagnetic needle kept at a fixed temperature $T=0.8T_c^p$ in the presence of an external magnetic field $h_{ext}=-0.25$. \\
  Bottom: Densities of spin $S_i^z = $+1, 0 and -1 ($\rho_{1}$, $\rho_{0}$ and $\rho_{-1}$) evolve with time. The vertical line in each figure indicates the reversal time for a single sample only. } 
  \label{fig:spin-den-pd}
\end{figure}

\begin{figure}[h!tpb]
 \centering
  \includegraphics[angle=0, width=1.00\textwidth]{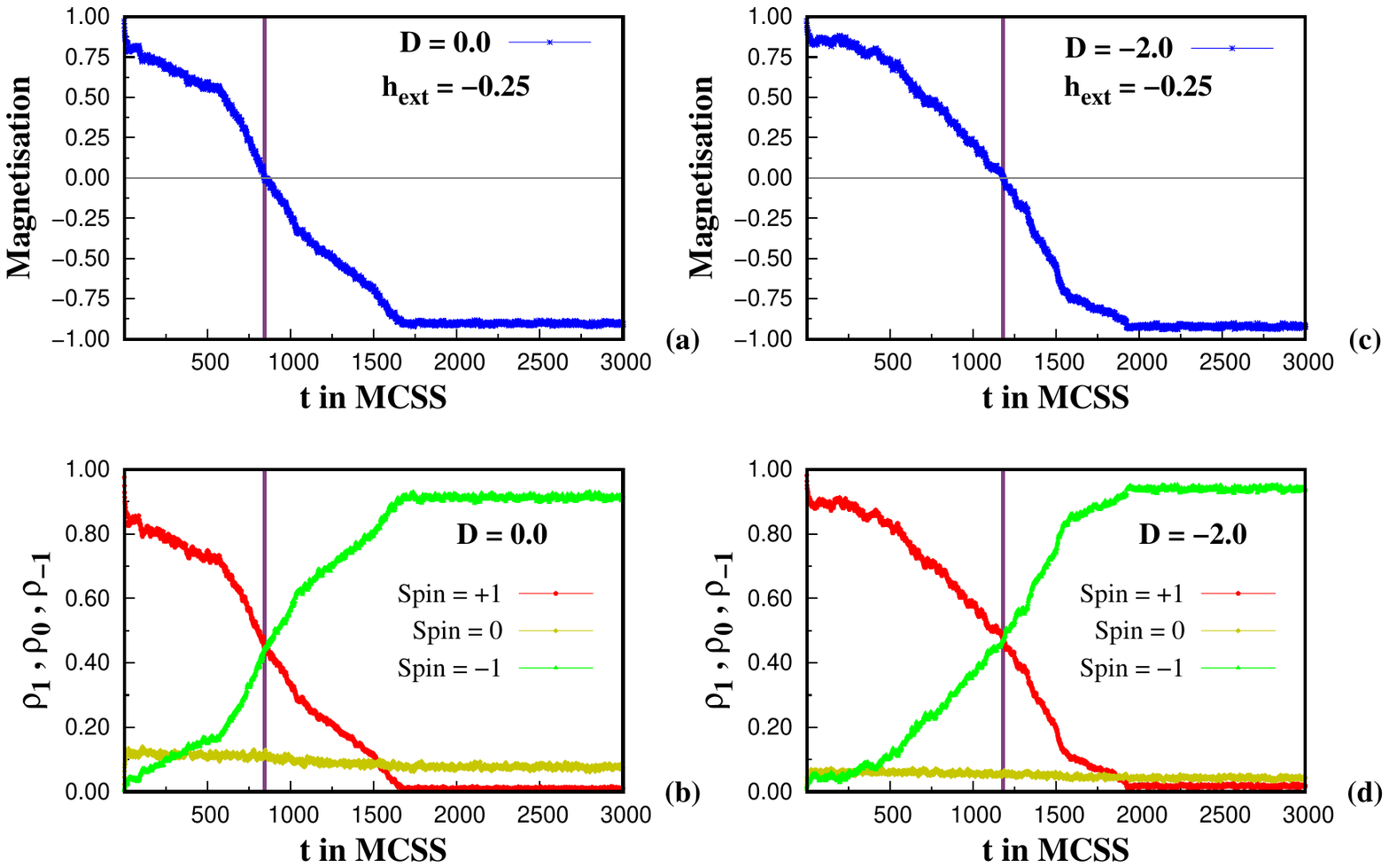}
  \caption{Top: Variation of magnetisation with time (in unit of MCSS) for $D=0$ and negative anisotropy of $D=-2.00$. The BC ferromagnetic needle kept at a fixed temperature $T=0.8T_c^p(D)$ in the presence of an external magnetic field $h_{ext}=-0.25$. \\
  Bottom: Corresponding time evolution of the densities of spin $S_i^z = $+1, 0 and -1 ($\rho_{1}$, $\rho_{0}$ and $\rho_{-1}$) for $D= 0~and~-2.00$. The vertical line in each figure indicates the reversal time for a single sample only.}
  \label{fig:spin-den-nd}
\end{figure}

\begin{figure}[h!tpb]
 \centering
  \includegraphics[angle=0, width=1.00\textwidth]{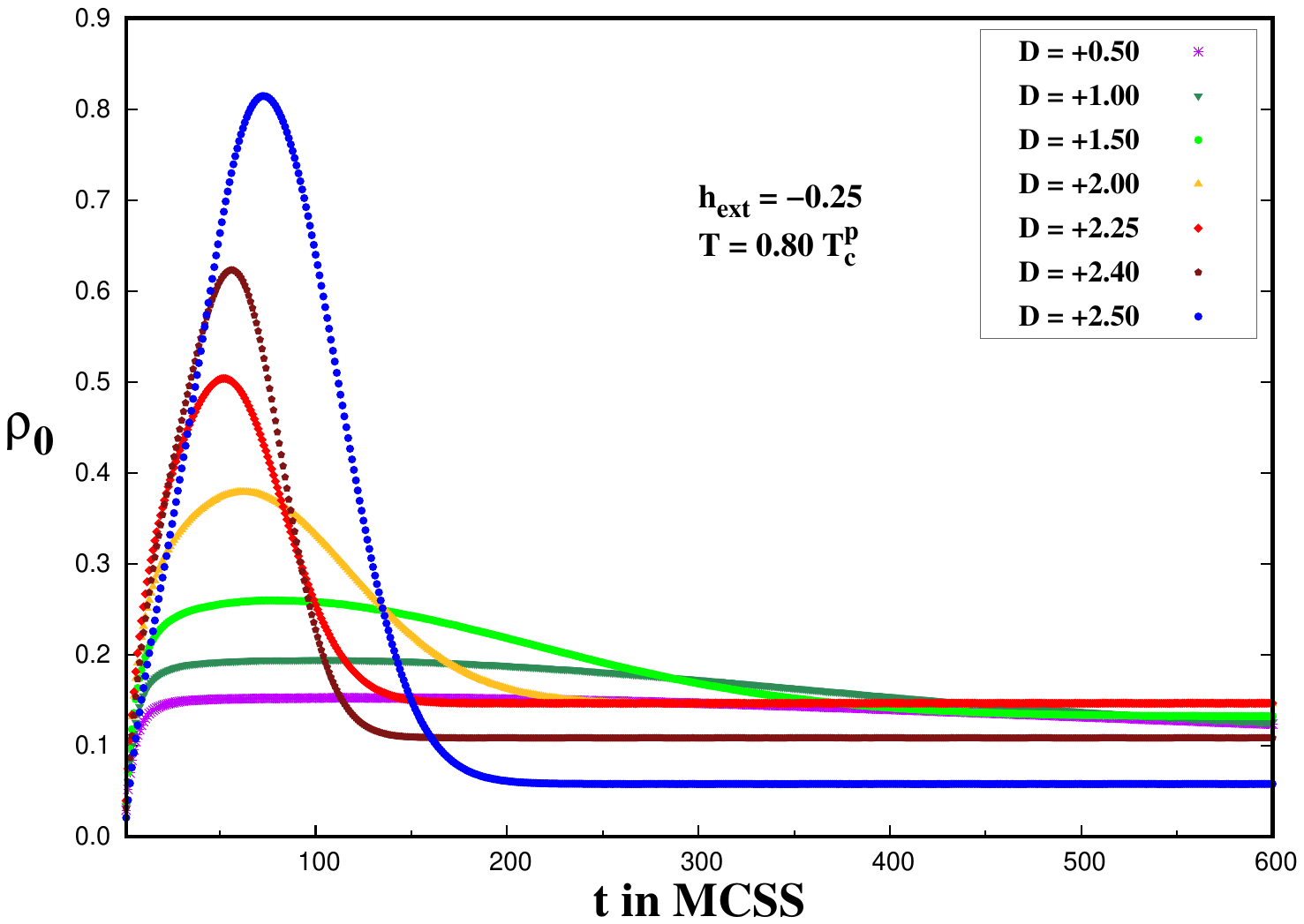}
  \caption{Temporal evolution of mean density of spin $S_i^z=0$, i.e. $\rho_0$ in a Blume-Capel ferromagnetic needle, illustrated for positive values of anisotropy parameter in the range of $0 < D \leq 2.50$. The system is maintained at a fixed temperature $T=0.8T_c^p(D)$ and external field $h_{ext} = -0.25$. Data are averaged over 10000 random samples.}
  \label{fig:spin0-den-pD}
\end{figure}

In this study, we have considered a needle-like Blume-Capel ferromagnetic system initially configured with the $z$ component of the spin at all lattice sites aligned upward $S_i^z = +1, \forall i$. The system is maintained in the ferromagnetic phase at a temperature $T=0.8T_c^p(D)$. Now, a weak magnetic field is applied in the negative $z$-direction. The system persists in a metastable state of positive magnetisation for a certain time period, even under the influence of a negative external magnetic field, and finally transitions to a stable state of negative magnetisation. The decay of magnetisation $m(t)$ with time is investigated, together with the dynamics of the densities of the spin components $S_i^z \in  \{+1,0,-1 \}$. The minimum time required by the system to overcome the metastable state and attain the negative magnetisation is referred to as the metastable lifetime or the reversal time $\tau_{rev}$ of magnetisation. Quantitatively, the reversal time $\tau_{rev}$ is defined as the minimum time steps (in MCSS) by which the magnetisation drops below $m=0$ and changes sign. Figures \ref{fig:spin-den-pd} and \ref{fig:spin-den-nd} represent the time evolution of magnetisation and the densities of spin $S_i^z = $+1, 0 and -1 (denoted by $\rho_{1}$, $\rho_{0}$ and $\rho_{-1}$ respectively) for different values of the anisotropy parameter $D$, where a Blume-Capel ferromagnetic needle is subjected to an external magnetic field $h_{ext} = -0.25$. The density of $S_i^z=+1$ ($\rho_{1}$) decays with time until it vanishes, whereas $\rho_{-1}$, starting from an initially negligible value, grows with time and eventually reaches its maximum. At the time of reversal, the densities of $S_i^z=+1$ and $S_i^z=-1$ become equal. It is observed from Fig. \ref{fig:spin-den-nd} that $\rho_0$, the density of $S_i^z=0$, remains vanishingly small throughout the entire time evolution in the presence of negative anisotropy $D<0$. On the other hand, for positive anisotropy, the $S_i^z=0$ is the most favourable state for minimizing the energy. Consequently, as positive anisotropy ($D>0$) increases, $\rho_0$, the density of $S_i^z=0$ maximises in the vicinity of the time of reversal (for moderate and higher values of positive anisotropy), as shown in Fig. \ref{fig:spin-den-pd}. Figure \ref{fig:spin0-den-pD} depicts how the mean density ($\rho_0$) of $S_i^z=0$, averaged over 10000 random samples of the Blume-Capel ferromagnetic needle, evolves with time for positive anisotropy values $0<D\le2.50$. The mean $\rho_0$ exhibits a pronounced peak near the reversal time, with the peak height increasing as the positive anisotropy becomes stronger. This is because under the influence of moderate and stronger positive anisotropy ($D>0$), the state $S_i^z=0$ becomes energetically more favourable, promoting its higher occupancy during magnetisation reversal.
\subsubsection{Metastable volume fraction:}
The study of the temporal decay of the metastable volume fraction is essential for obtaining clear insight into the spin dynamics and the nucleation mechanism in a needle-like Blume-Capel ferromagnetic system. Kolmogorov-Johnson-Mehl–Avrami (KJMA) \cite{K, JM, A1, A2, A3}  formalism, commonly referred to as the Avrami equation, is widely used to describe the kinetics of phase transformations involving the nucleation mechanism. The Avrami model for isothermal phase transformation with constant nucleation rate is described by \cite{avrami},
\begin{equation}
\label{avrami-eqn}
    \alpha(t)=1-\exp(-Kt^n)
\end{equation}
where $\alpha(t)$ is the time dependence of the transformed volume fraction. $K$ is the Avrami rate constant, depending on the temperature. `$n$' denotes the Avrami exponent, dependent of \emph{dimensionality $d$} of the system . The parameters $K$ and $n$ can be estimated from the following equation \cite{jabbar}, derived from the previous equation (\ref{avrami-eqn}) 
\begin{equation}
  \label{avrami_r} 
  \ln[-\ln(1-\alpha(t))] = n\ln(t) + \ln K 
\end{equation}
According to the Avrami model with interface-controlled constant nucleation rate $n=d+1$, where $d$ is the \emph{dimension of growth}.

In our study, the system is initially prepared with all spins aligned upward $S_i^z=+1, \forall i$. Under the influence of a negative external magnetic field, the relative abundance of up spins or the fraction of lattice sites with $S_i^z=+1$ decreases with time, characterising the decay of the metastable phase toward the stable down-spin configuration $S_i^z=-1$. This relative abundance of up spins is defined as metastable volume fraction $\beta=\frac{N_{+1}}{N}$. Previous Monte Carlo studies on metastability in the Blume-Capel ferromagnet \cite{NA, NA2, NAVF} have revealed that the decay of the metastable volume fraction follows Avrami's law. This naturally raises the question: Does the decay of metastable volume fraction in the BC ferromagnetic needle conform to the classical Avrami model, and how does the single-site anisotropy influence this metastable behaviour?

\begin{figure}[h!tpb]
 \centering
  \includegraphics[angle=0, width=1.00\textwidth]{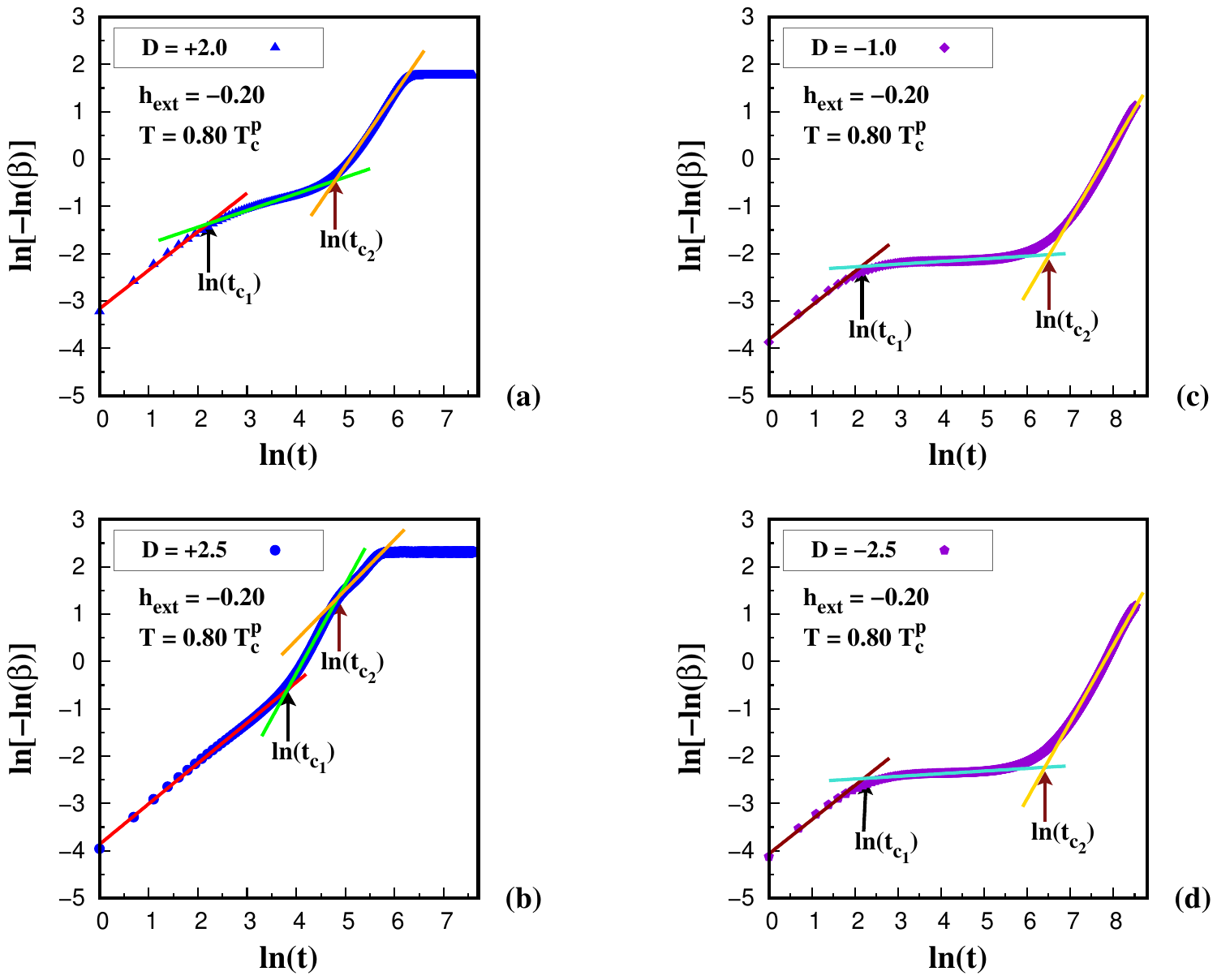}
  \caption{Time evolution of metastable volume fraction ($\beta = \frac{N_{+1}}{N}$) of spin component $S_i^z = +1$ in presence of external field $h_{ext}=-0.20$. The quantity $\ln{[-\ln{(\beta)}]}$ is plotted against $\ln(t)$ for four different values of anisotropy (a) $D=+2.00$, (b) $D=+2.50$, (c) $D=-1.00$ and $D=-2.50$. In each figure data are fitted to linear function $f(x)={a_1}x + a_0$ (where $f(x)= \ln[- \ln(\beta)]$ and $x= \ln(t) $) separately in three regimes, having different slope ($a_1$) value as indicated by solid straight lines. The crossover times are marked as $t_{c_{1}}$ and $t_{c_{2}}$, provided in Table \ref{tab:table1}.}
  \label{fig:v-frac-spinp1-20}
\end{figure}

\begin{figure}[h!tpb]
 \centering
  \includegraphics[angle=0, width=1.00\textwidth]{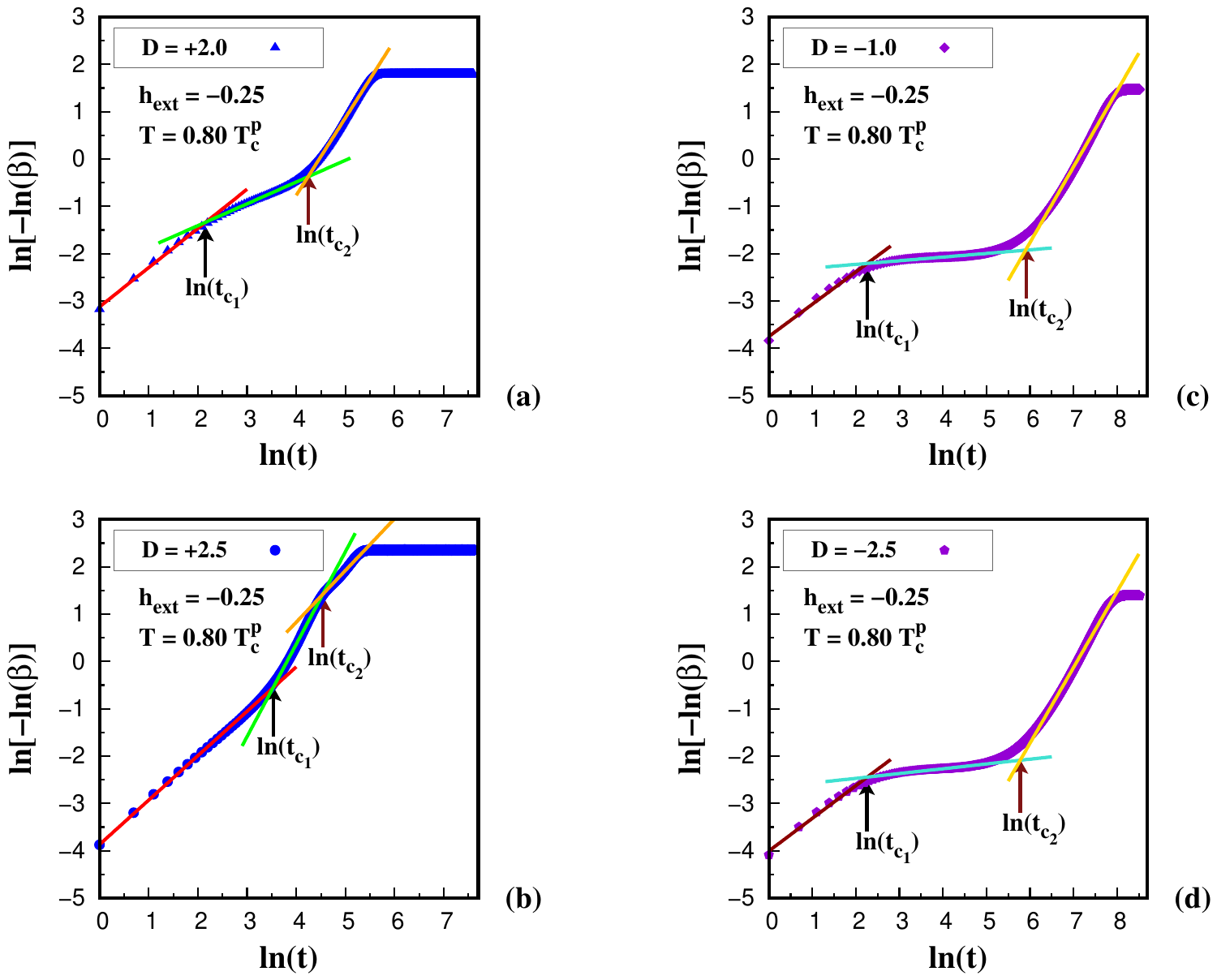}
  \caption{Similar as Fig. \ref{fig:v-frac-spinp1-20} for external field $h_{ext}=-0.25$. Time evolution of metastable volume fraction ($\beta = \frac{N_{+1}}{N}$) of spin component $S_i^z = +1$. The quantity $\ln{[-\ln{(\beta)}]}$ is plotted against $\ln(t)$ for four different values of anisotropy (a) $D=+2.00$, (b) $D=+2.50$, (c) $D=-1.00$ and $D=-2.50$. In each figure data are fitted to linear function $f(x)={a_1}x + a_0$ (where $f(x)= \ln[- \ln(\beta)]$ and $x= \ln(t) $) separately in three regimes, having different slope ($a_1$) value as indicated by solid straight lines. The crossover times are denoted as $t_{c_{1}}$ and $t_{c_{2}}$, provided in Table \ref{tab:table3}.}
  \label{fig:v-frac-spinp1-25}
\end{figure}

To address these questions, we have investigated $\ln[-\ln(\beta(t))]$ as a function of $\ln(t)$ for different values of the anisotropy parameter $D$ (both positive and negative), where the BC ferromagnetic needle is subjected to the negative external field of magnitude (i) $|h_{ext}|= 0.20$ (as shown in Fig. \ref{fig:v-frac-spinp1-20}) and (ii) $|h_{ext}|= 0.25$ (as represented in Fig. \ref{fig:v-frac-spinp1-25}). The temperature is maintained at $T=0.8T_c^p(D)$ (in the ferromagnetic phase). The data set obtained is statistically averaged over 10000 independent samples to ensure minimal fluctuation. Here, time ($t$) is expressed in dimensionless form by scaling with the unit MCSS. In the classical Avrami model, the plots $\ln[-\ln(\beta(t))]$ versus $\ln(t)$ are expected to yield a straight line, with the gradient corresponding to the Avrami exponent $n=d+1$. However, our findings reveal a notable deviation; instead of a single linear dependence, the curve can be partitioned into three distinct regimes, each of which is well described by a separate linear fit of the form $f(x)=a_1 x + a_0$, where $f(x) = \ln[- \ln(\beta)]$ and $x$ correspond to $\ln(t)$. The estimated slopes $a_1$ in these regimes quantitatively represent distinct effective exponents, thereby exposing clear crossovers between multiple dynamical regimes during metastable decay. The corresponding crossover times, $t_{c_1}$ and $t_{c_2}$, marked in Fig. \ref{fig:v-frac-spinp1-20} and Fig. \ref{fig:v-frac-spinp1-25} are systematically summarized in Table \ref{tab:table1} and Table \ref{tab:table3}. The best-fit parameters and the value of corresponding $\chi^2$ (and DOF) related to the linear fit $f(x)=a_1 x + a_0$ are provided in Table \ref{tab:table2} and Table \ref{tab:table4}. Such a multi-regime behaviour starkly contrasts with the single-exponent prediction of the classical Avrami model, highlighting that the decay of the metastable volume fraction in the needle-like Blume-Capel ferromagnet manifests distinctive dynamical features beyond the Avrami expectation.

\begin{figure}[h!tpb]
 \centering
  \includegraphics[angle=0, width=1.00\textwidth]{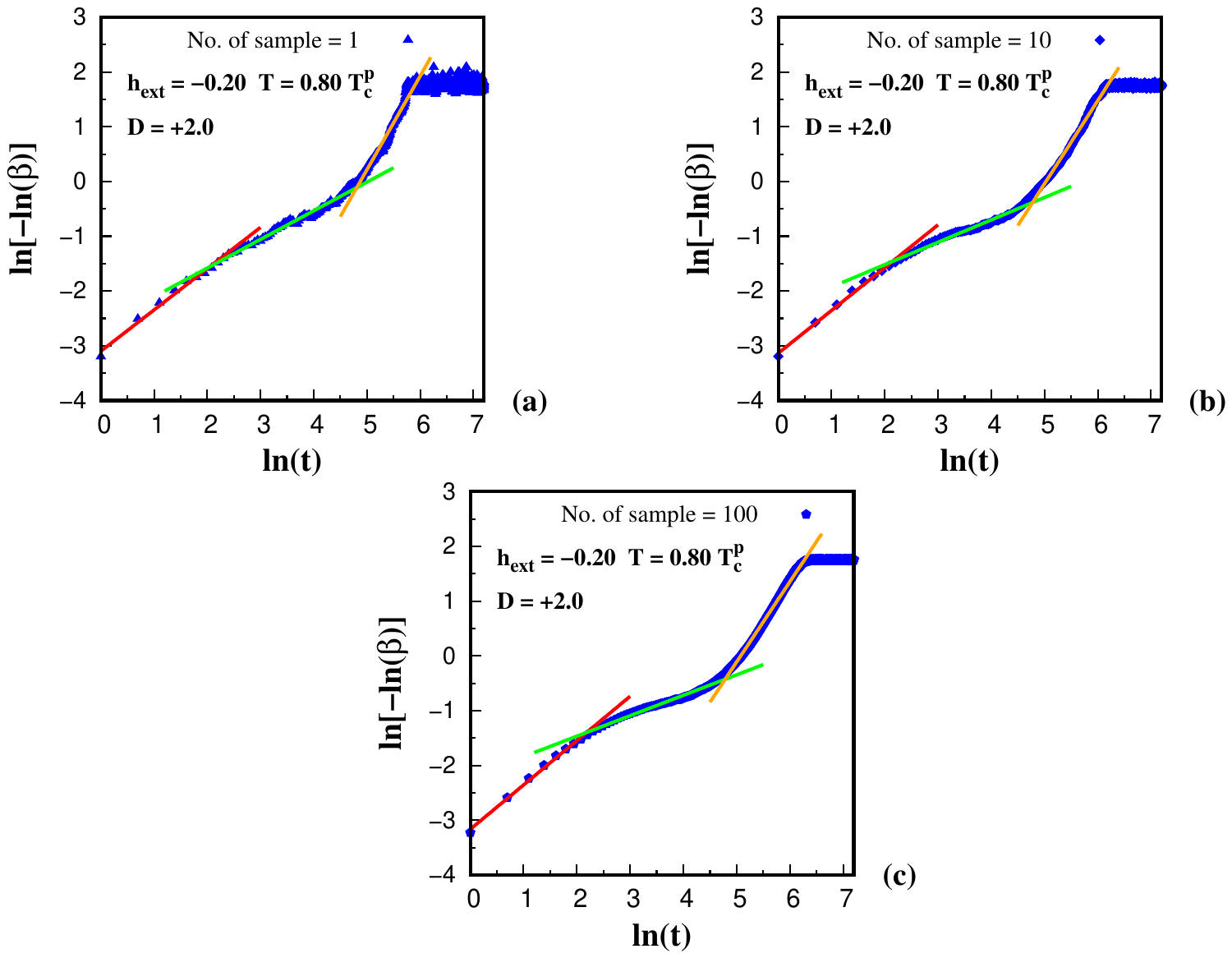}
  \caption{\textcolor{blue}{The decay of metastable volume fraction ($\beta = \frac{N_{+1}}{N}$) of spin component $S_i^z = +1$ under an external field $h_{ext}=-0.20$ at a fixed value of anisotropy $D=+2.0$ for a Blume–Capel needle of size $L_x=L_y=4$ and $L_z=200$. The plots display $\ln{[-\ln{(\beta)}]}$ versus $\ln(t)$, where the curves correspond to averages taken over (a) 1, (b) 10, and (c) 100 independent samples. In each case, the evolution is analysed across three distinct time intervals, with linear fits of the form $f(x)={a_1}x + a_0$, where $f(x)= \ln[- \ln(\beta)]$ and $x= \ln(t)$. Solid straight lines denote the fitted segments for each regime. The fitting statistics are provided in Table \ref{tab:sample}.}}
  \label{fig:v-frac-sample}
\end{figure}

\textcolor{blue}{It may be noted that very small values have been obtained for $\chi^2$ with various values of corresponding DOF. However, if one check with the hypothesis testing (considering the null hypothesis
 as $H_0:$ the data fits with exponential, and alternate hypothesis $H_1:$ the alternate fitting), the observed $\chi^2$ values are 
well below the significant level $\chi_{\alpha}$ above which the distributions (probability distribution function for $k$ degrees of freedom) of $\chi^2$, i.e. $P_{k}(x)={{1} \over {2^{k/2} {\Gamma(k/2)}}}x^{(k/2)-1} e^{-x/2}$,  having area five percent of total area. We have checked it for $D=+2.0$, $h_{ext}=-0.20$ and the first region for DOF=7, the $\chi^2=0.0125$
(see first line of Table~\ref{tab:table2}), which is much less than 
$\chi^2_{\alpha}$= 14.04(from $\chi^2$ table). In all other cases, the value of $\chi^2$ is much smaller than the value of
corresponding significant level $\chi^2_{\alpha}$.
This
suggests that there is no reason to reject null hypothesis for this case. So, the exponential fitting of the decay of metastable volume fraction
is reasonably accepted. Our results of all $\chi^2$ test of the fitting clearly suggests the acceptance of null hypothesis. The small values of $\chi^2$ arises from the the data of metastable volume fraction averaged over large number of samples. To confirm this we have studied the behaviour for various number of samples (1, 10, 100 here) as shown in Figure \ref{fig:v-frac-sample}. The results indicate that, for smaller sample sizes, the fluctuations are significantly larger, leading to correspondingly higher $\chi^2$ in the fitting. The fitting statistics is provided in Table \ref{tab:sample} for different number of samples.}

\begin{figure}[h!tpb]
 \centering
  \includegraphics[angle=0, width=1.00\textwidth]{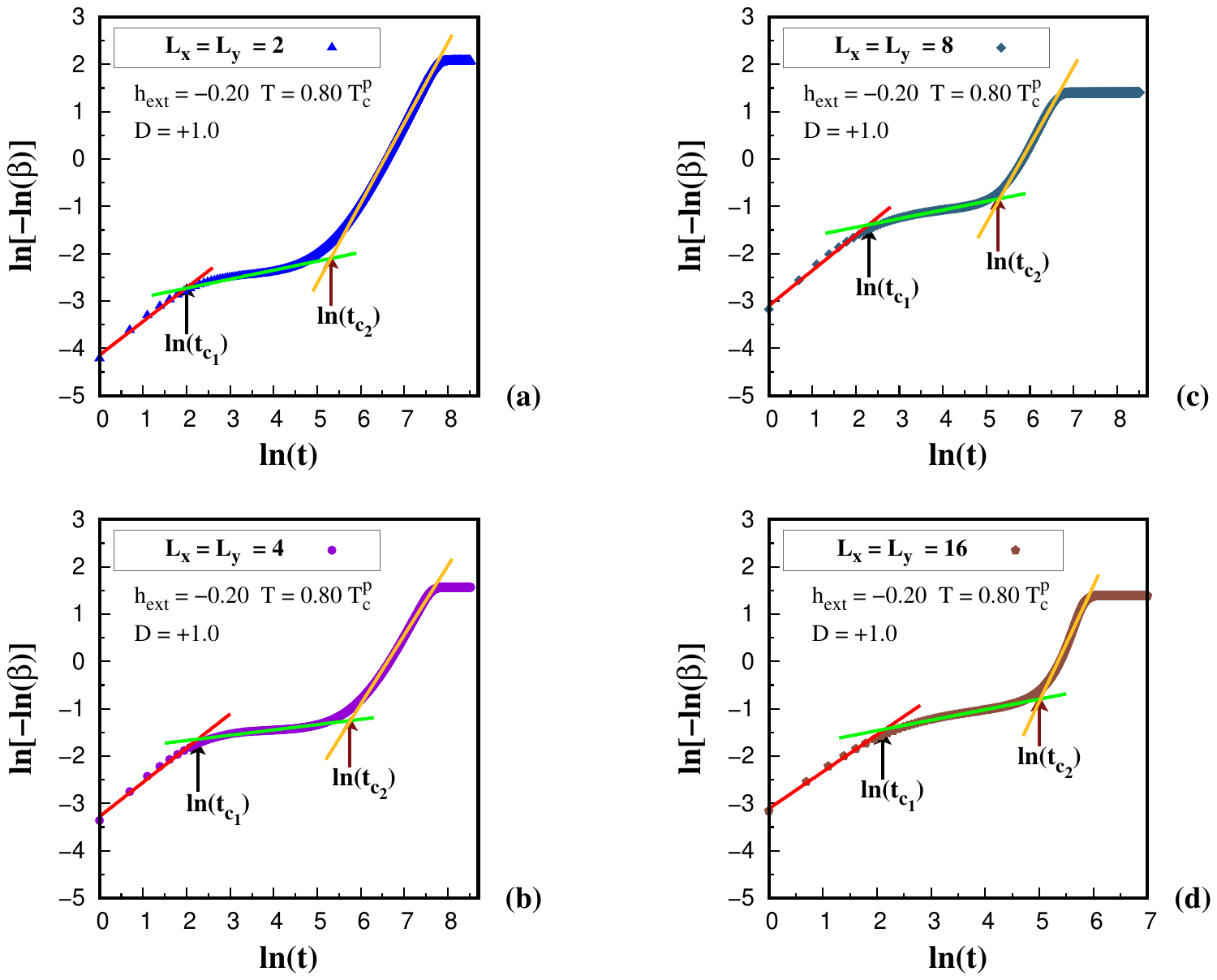}
  \caption{\textcolor{blue}{Decay of metastable volume fraction ($\beta = \frac{N_{+1}}{N}$) of spin component $S_i^z = +1$ as a function of time under an external field $h_{ext}=-0.20$ for different values of $L_x(=L_y)$. The quantity $\ln{[-\ln{(\beta)}]}$ is plotted against $\ln(t)$ for a fixed value of anisotropy $D=+1.0$ , while varying the cross-sectional dimensions ($L_x=L_y$) of the Blume-Capel ferromagnetic needle and keeping the longitudinal size fixed at $L_z=200$. In each panel, the data are fitted using a linear function $f(x)={a_1}x + a_0$, where $f(x)= \ln[- \ln(\beta)]$ and $x= \ln(t) $, separately over three distinct time regimes. The corresponding slopes $a_1$ for each regime are shown with solid straight lines. The crossover times are marked as $t_{c_{1}}$ and $t_{c_{2}}$are indicated on the plots and listed in Table \ref{tab:crosstime}.}}
  \label{fig:v-frac-Lxy}
\end{figure}

\textcolor{blue}{In the above discussion, the decay of metastable volume fraction, is shown for a single value of the cross -section of the BC needle. Here, $L_x=L_y=4$. A relevant question may arise here, how does this behaviour depends on the
cross sectional area of the needle. To address this issue, we have investigated the decay of metastable volume fraction for three more values of the
cross-sectional area of the needle, namely $L_x=L_y=2$, 8 and 16. The results are shown in Figure \ref{fig:v-frac-Lxy} The same kind of exponential decay is
observed, while exhibiting three clearly distinguishable dynamical regimes. Which reveals the multiple crossover of the deacy of the metastable volume fraction. However, the values of $n$ are found
to depend on the cross-sectional area. The values of $n$ and the crossover time for different cross sectional area and are
provided in a Table \ref{tab:crosstime}}.



\subsubsection{Effect of anisotropy on Reversal time:}
\begin{figure}[h!tpb]
 \centering
  \includegraphics[angle=0, width=1.00\textwidth]{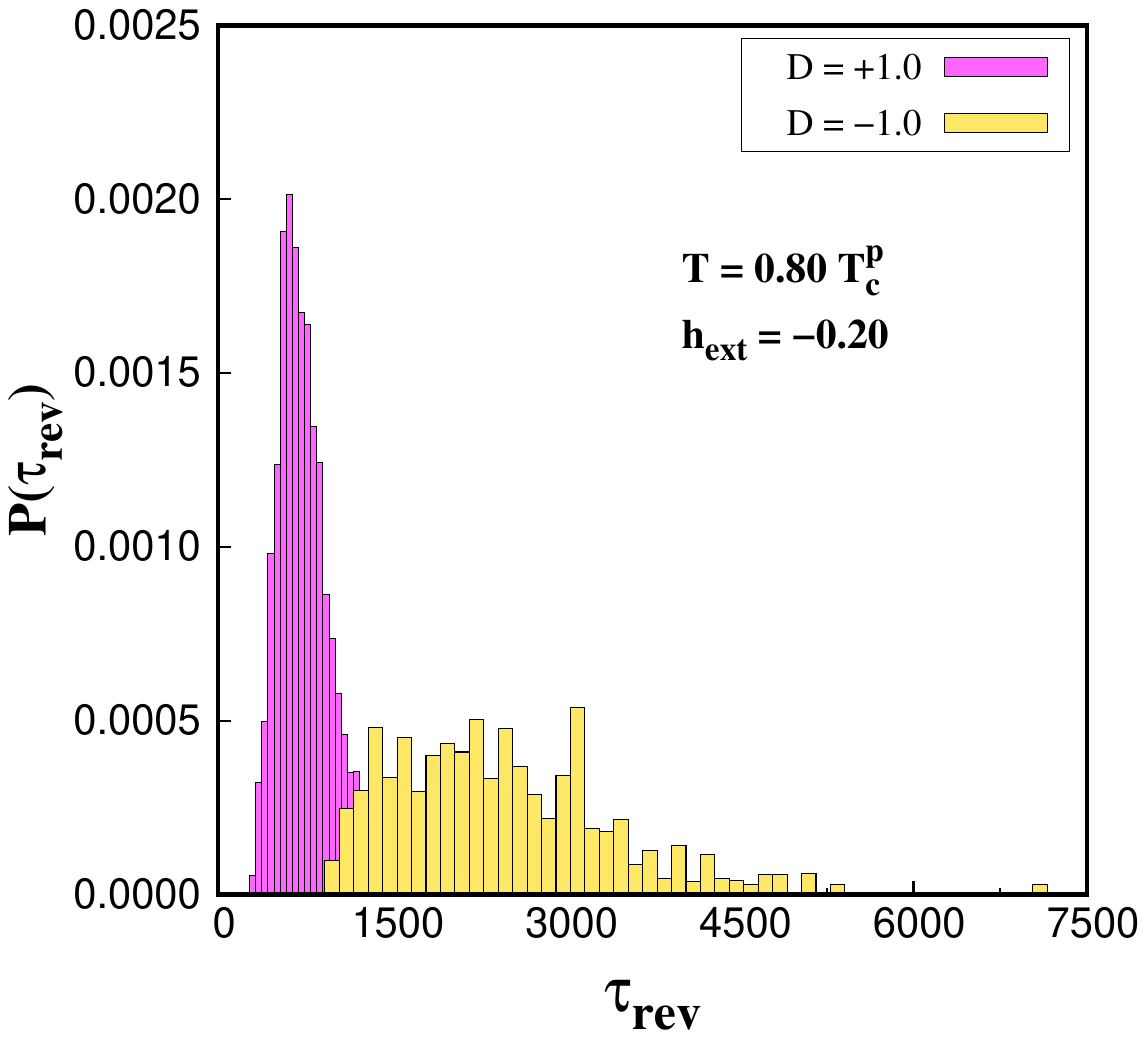}
  \caption{Normalised probability distribution ($P (\tau_{rev})$) of the reversal times($\tau_{rev}$) for same absolute value of positive and negative anisotropy $D=+1.0~and~-1.0$. Temperature of the system fixed at $T=0.8T_c^p(D)$ and external field $h_{ext}=-0.20$. Data are obtained for 10000 random samples.}
  \label{fig:dist-rev-time}
\end{figure}

The reversal time $\tau_{rev}$ of the BC ferromagnetic needle has been systematically analysed to elucidate the influence of single-site uniaxial anisotropy. To ensure statistical reliability, we have calculated the mean value of $\tau_{rev}$ averaging over 10000 random samples. The statistical distribution of the reversal times is depicted in Fig. \ref{fig:dist-rev-time} for the same strength of positive and negative anisotropy, $D=+1.0$ and $D=-1.0$. These results are obtained for a system under $h_{ext}=-0.20$, kept at a fixed temperature $T=0.8T_c^p(D)$ (in the ferromagnetic phase). The normalised probability distribution of $\tau_{rev}$ exhibits significantly larger fluctuations in the presence of negative anisotropy ($D<0$), resulting in a considerably wider distribution. On the other hand, the distribution of $\tau_{rev}$ under positive anisotropy ($D>0$) is much narrower than that for negative anisotropy, and the spread of the distribution further decreases as the positive anisotropy strength increases. 

\begin{figure}[h!tpb]
 \centering
  \includegraphics[angle=0, width=1.00\textwidth]{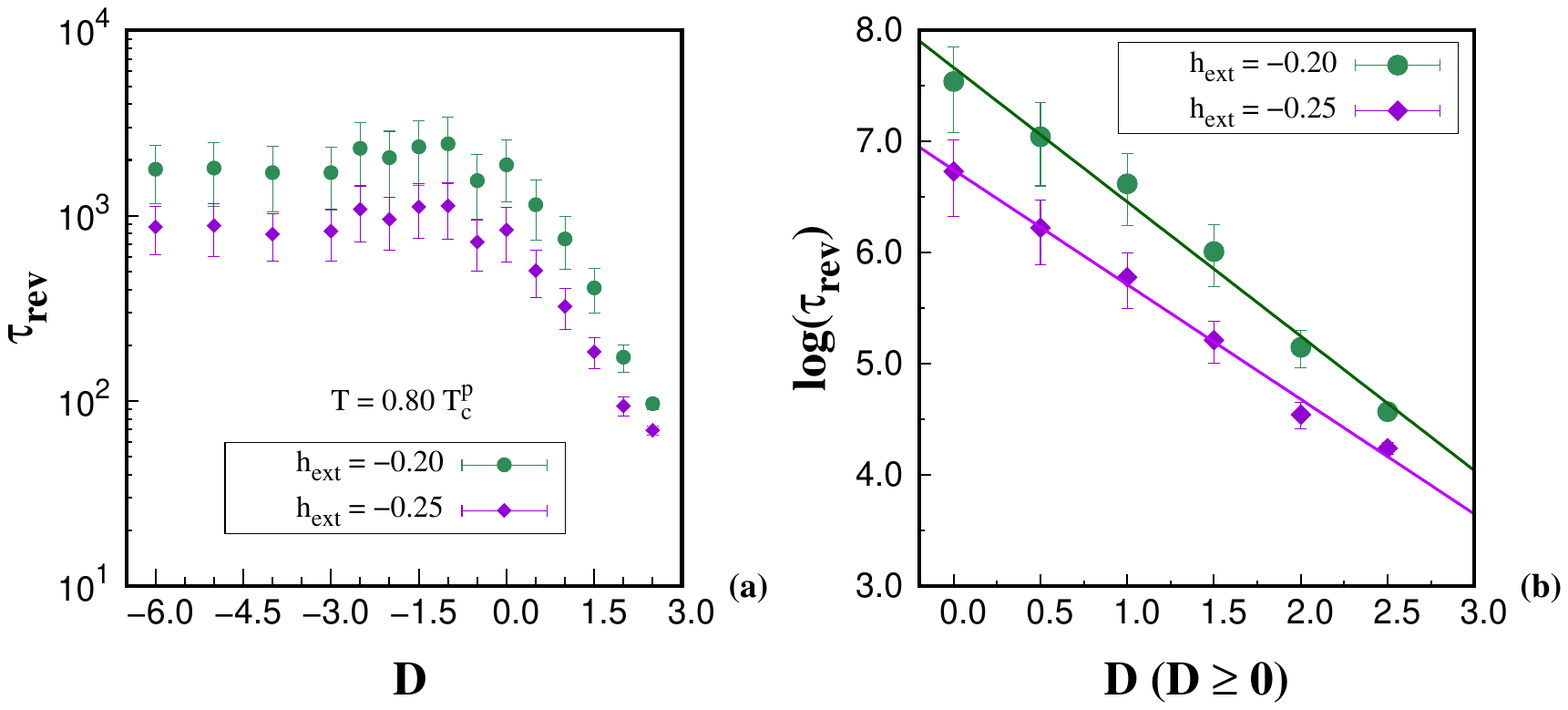}
  \caption{(a) Mean reversal time $\tau_{rev}$ as a function of anisotropy parameter $D$, plotted in semi-logarithmic scale, obtained for 10000 random samples of Blume-Capel ferromagnetic needle. The green circle symbolises the data for $h_{ext}=-0.20$ and the violet diamond for $h_{ext}=-0.25$. The temperature is fixed at $T=0.8T_c^p(D)$. \\
  (b) Logarithmic values of mean reversal time $\log(\tau_{rev})$ are plotted against positive anisotropy ($0 \le D \le 2.5$). Data are fitted to a linear function $g(x)= b_0 + {b_{1}}x$, where $g(x)$ stands for $\log(\tau_{rev})$ and $x=D$ (for $D \ge 0$). The green solid line fitted for $h_{ext}=-0.20$ and violet for $h_{ext}=-0.25$. The values of $b_1$ are provided in the Table \ref{tab:table5}.}
  \label{fig:reversaltime}
\end{figure}

\begin{figure}[h!tpb]
 \centering
  \includegraphics[angle=0, width=1.00\textwidth]{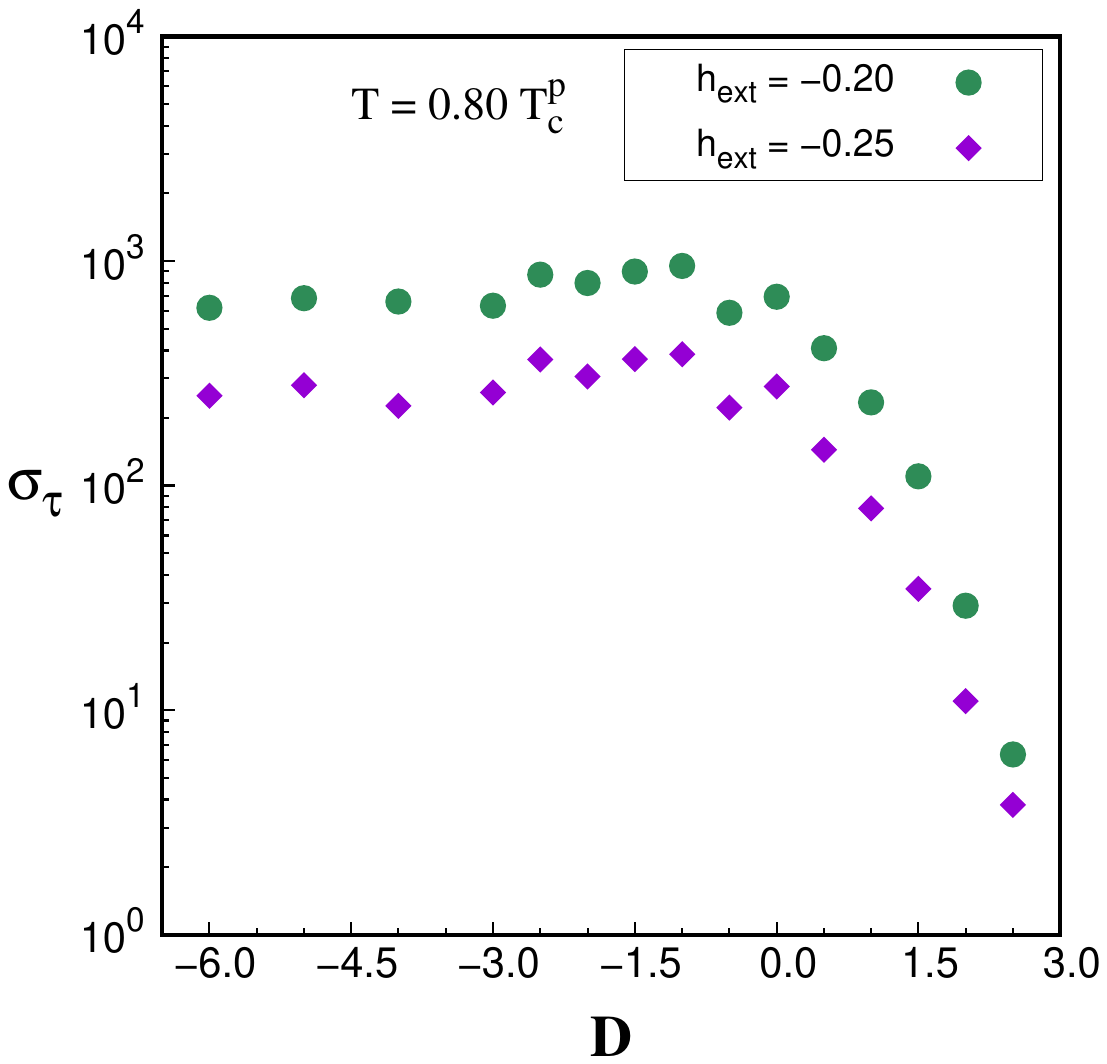}
  \caption{The standard deviation ($\sigma_{\tau}$) of reversal times on logarithmic scale plotted against the anisotropy parameter $D$. The green circle symbolises the data for $h_{ext}=-0.20$ and the violet diamond for $h_{ext}=-0.25$. Results are obtained for 10000 random samples of the BC ferromagnetic needle. Temperature kept fixed at $T=0.8T_c^p(D)$.}
  \label{fig:sdrev}
\end{figure}

We have studied the mean reversal time $\tau_{rev}$ of the BC ferromagnetic needle as a function of anisotropy $D$ under the external magnetic field values $h_{\text{ext}}=-0.20$ and $h_{\text{ext}}=-0.25$, as represented in Fig. \ref{fig:reversaltime}. Figure \ref{fig:reversaltime} (b) demonstrates that the logarithmic value of the mean reversal time $\log(\tau_{rev})$ linearly decreases with increasing strength $|D|$ under the influence of positive anisotropy ($D \ge 0$). Reversal time data under positive anisotropy are well represented by a linear fit of the form $g(x)=b_0 + b_1 x$, where $g(x)$ corresponds to $\log(\tau_{rev})$ and $x$ denotes $D$ for $D > 0$. This implies an exponential decrease of $\tau_{rev}$ with increasing positive anisotropy; $\tau_{rev} \sim e^{-b_1 D}$. The best fit parameters ($b_1 \pm \delta b_1$) together with statistical $\chi^2$ and degrees of freedom (DoF) are listed in Table \ref{tab:table5}. In contrast, the effect of negative anisotropy on the mean reversal time $\tau_{rev}$ is comparatively weak and insignificant ($\tau_{rev}$ is almost independent of negative $D$), as depicted in Fig. \ref{fig:reversaltime} (a). However, strong positive anisotropy results in significantly shorter reversal times compared to the negative anisotropy case. The reason behind this scenario can be explained by the Blume-Capel model. The single-site anisotropy term in the BC Hamiltonian (equation \ref{hamiltonian}) $D\sum_{i}{(S_i^z)}^2$ contributes the energetic preferences among the spin states $S_i^z \in \{+1, 0, -1\}$. The positive anisotropy ($D>0$) favours the spin state $S_i^z=0$ to lower the energy, leading to enhanced population of $S_i^z=0$. The large production of non-magnetic sites or vacancies effectively reduces the stability of the metastable state. That accelerates the magnetisation reversal and shortens the reversal time. On the contrary, for the negative $D$, the states $S_i^z= \pm 1$ are energetically more preferable, strengthening the magnetisation. That results in prolonged metastability under negative anisotropy. The anisotropy dependence of the standard deviation of reversal times $\sigma_{\tau}$ is also studied and illustrated in Fig. \ref{fig:sdrev}. The standard deviation $\sigma_{\tau}$ is found to decrease with increasing positive anisotropy $D>0$, whereas $\sigma_{\tau}$ is almost insensitive to negative anisotropy $D<0$. 

\subsection{Magnetic relaxation in BC ferromagnetic needle:}
\begin{figure}[h!tpb]
 \centering
  \includegraphics[angle=0, width=1.00\textwidth]{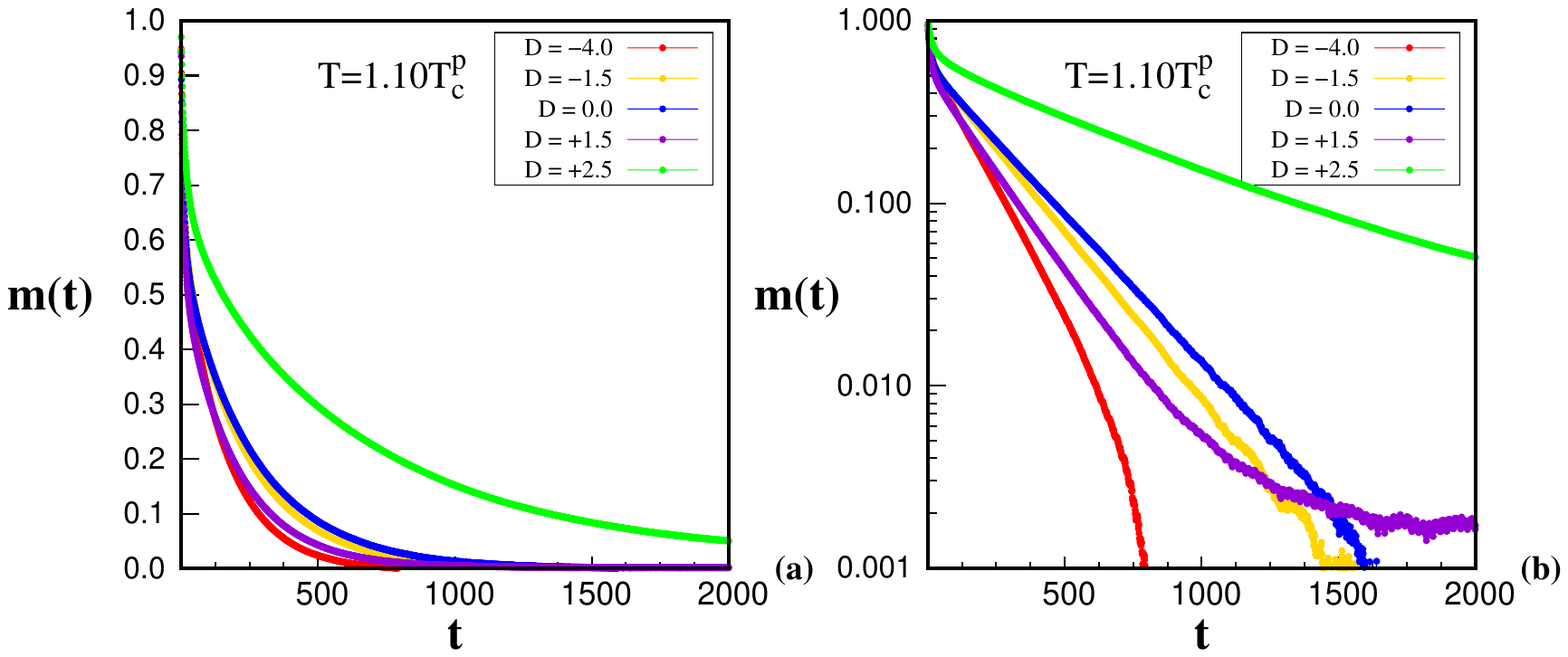}
  \caption{The temporal decay of magnetisation $m(t)$ for different values of anisotropy $D$ plotted in (a) linear scale and (b) semi-logarithmic scale. Data obtained by averaging over 20000 random samples of the BC ferromagnetic needle. Temperature maintained at $T=1.10T_c^p(D)$.}
  \label{fig:m-relaxation}
\end{figure}

We next turn our attention to the magnetic relaxation in the needle-shaped Blume-Capel ferromagnetic system. In order to study magnetic relaxation, we initially configured the system with all spins pointed up, $S_i^z=+1, \forall i$, establishing the initial state of magnetisation $m=1$ and then, in the absence of any external magnetic field, let the system relax to its spontaneous magnetisation at a temperature above the critical point $T_c$. The magnetisation of the system decays with time (in unit of MCSS) and eventually vanishes as time goes to infinity. In our simulation, we have maintained the equivalent thermal condition for the system with different values of the anisotropy parameter $D$ by fixing the temperature $T=f \times T_c^p(D)$, where $f= 1.10$ is considered to ensure that the system is in the paramagnetic phase. The time evolution of magnetisation, averaged over 20000 random samples of the BC ferromagnetic needle, is systematically examined under positive and negative values of the anisotropy parameter $D$. The relaxation of magnetisation is observed to follow an exponential decay for any value of single-site anisotropy $D$, as shown in Fig. \ref{fig:m-relaxation} (a) in linear scale and (b) in semi-logarithmic scale. \[ m(t) \sim \exp (-t/\tau_{relax}) \] where $\tau_{relax}$ is the characteristic time scale referred to as \emph{relaxation time}. The relaxation time can be extracted from the slope of the semi-log plot of $m(t)$ fitted to an exponential function $f(t)=a\exp(- bt/10^{3})$. However, we could not extract any functional dependence of the relaxation time on the anisotropy. 
\\


\section{Concluding remarks:}
In this manuscript, we mainly report our Monte Carlo results of the transient behaviours of a Blume-Capel (BC) ferromagnetic needle. The ferromagnetic needle has a much longer elongated length than its cross-sectional area. We have studied the transient behaviours under the influence of single-site uniaxial anisotropy. Firstly, we have secured a rough estimate of the anisotropy-dependent critical temperature. This is qualitatively similar to the previous studies of anisotropy-dependent critical temperature for the bulk BC ferromagnet. These results are crucial in ensuring an equivalent thermal state of the system for different values of anisotropy. \textcolor{blue}{As an extension, the dependence of pseudocritical temperature on the cross sectional area of the Blume-Capel needle has been studied. The pseudocritical temperature has been found to increase. The
value of the pseudocritical temperature for bulk cubic sample is believed to be achieved\cite{ozkan,kutlu} as the cross sectional area increases.}

Within the broader context of transient behaviour, we have investigated the temporal decay of the metastable volume fraction ($\beta$). This is usually analysed in the light of Avrami's law $\beta(t) \sim {\rm exp}(-Kt^n)$ in the bulk ferromagnet. Previous studies on bulk samples \cite{surface-bulk, NA} show that the system usually obeys Avrami's law, where $n=d+1$ and $d$ is the dimensionality of the system, where nucleation occurs. However, in our present study, we have observed the unusual deviation of Avrami's exponent ($n\neq {d+1}$) for the first time. Not only shows the simple deviation from Avrami's law, the Blume-Capel ferromagnetic needle also exhibits multiple crossovers (multiple distinct values of $n$) in the decay of the metastable volume fraction. This is an interesting result which has not yet been reported anywhere. \textcolor{blue}{The decay of metastable 
volume fraction has been investigated for different cross-sectional area of the needle. The $n$ and the crossover time have been found to depend on the
cross sectional area}.

In this context, the magnetisation reversal time has been studied as a function of the strength of single-site anisotropy. The mean reversal time is found to be almost independent of the anisotropy for its negative values. However, for positive values of the anisotropy, the mean reversal time has been found to decrease exponentially. The distribution of reversal times exhibits a broader dispersion under negative anisotropy compared to the narrow distributions observed for moderate and strong positive anisotropy.  

The magnetic relaxation behaviour (in the paramagnetic phase) has been found to be exponential, as generally shown by the bulk sample. However, our results are inconclusive to determine any functional dependence of the relaxation time on the strength of the anisotropy. 

\vskip 1cm
\noindent {\bf Acknowledgements:}  
IT acknowledges UGC JRF, Govt. of India, for financial support. We are thankful to the computational facilities provided by Presidency University, Kolkata. 
\vskip 0.2cm
\noindent {\bf Data availability statement:} Data may be available on reasonable request to Ishita Tikader.

\vskip 0.2cm

\noindent {\bf Code availability statement:} Code may be available on reasonable request to Ishita Tikader.

\vskip 0.2cm

\noindent {\bf Conflict of interest statement:} We declare that this manuscript is free from any conflict of interest.
\vskip 0.2cm

\noindent {\bf Funding statement:} No funding was received, particularly to support this work.

\vskip 0.2cm

\noindent {\bf Authors’ contributions:} Ishita Tikader developed the code, prepared the figures, and wrote the manuscript.
Muktish Acharyya conceptualised the problem, analysed the results and wrote the manuscript.

\newpage
\newpage
\newpage
\newpage

\newpage

\newpage
\newpage

\newpage
\newpage

\newpage
\newpage

\newpage

\newpage

\newpage
\newpage
\newpage
\begin{table*}[ht!]
\small
  \caption{Data of crossover times corresponding to Fig.\ref{fig:v-frac-spinp1-20} for external field $h_{ext}=-0.20$.}
  \label{tab:table1}
  \begin{tabular*}{\textwidth}{@{\extracolsep{\fill}}ccccc}
  \hline
  Anisotropy($D$) & $\ln(t_{c_{1}})$  & 1st Crossover time ($t_{c_{1}}$) & $\ln(t_{c_{2}})$ & 2nd Crossover time ($t_{c_{2}}$)  \\
 \hline
 +2.0 & 2.217 & Around 9  &  4.785 & Around 120   \\
 +2.5 & 3.823 & Around 46  &  4.862 & Around 130   \\
 -1.0 & 2.160 & Around 9 & 6.509 & Around 672 \\
 -2.0 & 2.200 & Around 9 & 6.415 & Around 611 \\
\hline
 \end{tabular*}
\end{table*}

\begin{table*}[ht!]
\small
  \caption{Fitting parameters corresponding to Fig.\ref{fig:v-frac-spinp1-20} for external field $h_{ext}=-0.20$.}
  \label{tab:table2}
  \begin{tabular*}{\textwidth}{@{\extracolsep{\fill}}ccccc}
  \hline
  Anisotropy ($D$) & Time regime & $n = a_1\pm \delta a_1$ & $\chi^2$ & DoF \\
 \hline
 
 \multirow{3}{*}{+2.0}
         & Region I & $0.814 \pm 0.021$  & 0.0125 & 7   \\
         & Region II & $0.352 \pm 0.005 $ & 0.0917 & 98 \\
         & Region III & $1.518 \pm 0.006 $ & 0.3494 & 280 \\
 \hline
 
 \multirow{3}{*}{+2.5}
         & Region I & $0.856 \pm 0.005 $ & 0.0189 & 31  \\
         & Region II & $1.890 \pm 0.007 $ & 0.0288 & 83  \\
         & Region III & $1.062 \pm 0.006$ & 0.08130 & 194  \\
 \hline
 
 \multirow{3}{*}{-1.0}
         & Region I & $0.717 \pm 0.030$ & 0.0129 & 5  \\
         & Region II & $0.057 \pm 0.001$ & 0.1007 & 317  \\
         & Region III & $1.547 \pm 0.002$ & 25.3940 & 4247 \\
  \hline 
  
  \multirow{3}{*}{-2.5}
         & Region I & $0.722 \pm 0.029 $ & 0.0122 & 5  \\
         & Region II & $0.056 \pm 0.001$ & 0.0378 & 231  \\
         & Region III & $1.617 \pm 0.002 $ & 17.5775 & 4247  \\
   \hline 
   
  \end{tabular*}
\end{table*}

\begin{table*}[ht!]
\small
  \caption{Data of crossover times corresponding to Fig.\ref{fig:v-frac-spinp1-25} for external field $h_{ext}=-0.25$.}
  \label{tab:table3}
  \begin{tabular*}{\textwidth}{@{\extracolsep{\fill}}ccccc}
  \hline
  Anisotropy($D$) & $\ln(t_{c_{1}})$  & 1st Crossover time ($t_{c_{1}}$) & $\ln(t_{c_{2}})$ & 2nd Crossover time ($t_{c_{2}}$)  \\
 \hline
 +2.0 & 2.150 & Around 9  &  4.240 & Around 70   \\
 +2.5 & 3.525 & Around 34  &  4.540 & Around 94   \\
 -1.0 & 2.260 & Around 10 & 5.920 & Around 372 \\
 -2.0 & 2.250 & Around 10 & 5.769 & Around 320 \\
\hline 
 \end{tabular*}
\end{table*}

\begin{table*}[ht!]
\small
  \caption{Fitting parameters corresponding to Fig. \ref{fig:v-frac-spinp1-25} for external field $h_{ext}=-0.25$}
  \label{tab:table4}
  \begin{tabular*}{\textwidth}{@{\extracolsep{\fill}}ccccc}
  \hline
  Anisotropy ($D$) & Time regime & $n = a_1\pm \delta a_1$ & $\chi^2$ & DoF \\
 \hline
 
 \multirow{3}{*}{+2.0}
         & Region I & $0.828 \pm 0.019 $  & 0.0107 & 7   \\
         & Region II & $ 0.461 \pm 0.004 $ & 0.0055 & 43 \\
         & Region III & $1.640 \pm 0.008 $ & 0.0874 & 125 \\
 \hline
 
 \multirow{3}{*}{+2.5}
         & Region I & $0.939 \pm 0.006 $ & 0.0255 & 31 \\
         & Region II & $1.955 \pm 0.016 $ & 0.0380 & 46 \\
         & Region III & $1.094 \pm 0.010$ & 0.2569 & 161  \\
 \hline
 
 \multirow{3}{*}{-1.0}
         & Region I & $0.683 \pm 0.030$ & 0.0269 & 7   \\
         & Region II & $0.077 \pm 0.002 $ & 0.0297 & 137 \\
         & Region III & $1.601 \pm 0.002 $ & 7.9400 & 2575  \\
  \hline
  
  \multirow{3}{*}{-2.5}
         & Region I & $0.691 \pm 0.029$ & 0.0248 & 7  \\
         & Region II & $0.101 \pm 0.002$ & 0.0345 & 137  \\
         & Region III & $1.600 \pm 0.002 $ & 5.6151 & 2575  \\
   \hline
   
   \end{tabular*}
\end{table*}

\begin{table*}[ht!]
\small
  \caption{\textcolor{blue}{Fitting parameters corresponding to Fig.\ref{fig:v-frac-sample} for external field $h_{ext}=-0.20$ and fixed anisotropy $D=+2.0$.}}
  \label{tab:sample}
  \begin{tabular*}{\textwidth}{@{\extracolsep{\fill}}ccccc}
  \hline
  No. of sample & Time regime & $n = a_1$ & $\chi^2$ & DoF \\
 \hline
 
 \multirow{3}{*}{1}
         & Region I & $ 0.753 $  & $ 0.014 $ & 7   \\
         & Region II & $ 0.523 $ & $ 0.182 $ & 19 \\
         & Region III & $ 1.711 $ & $ 0.670 $ & 26 \\
 \hline
 
 \multirow{3}{*}{10}
         & Region I & $ 0.781 $ & $ 0.007 $ & 7  \\
         & Region II & $ 0.409 $ & $ 0.078 $ & 19  \\
         & Region III & $ 1.519 $ & $ 0.248 $ & 26  \\
 \hline
 
 \multirow{3}{*}{100}
         & Region I & $ 0.809 $ & $ 0.007 $ & 7  \\
         & Region II & $ 0.373 $ & $ 0.041 $ & 19  \\
         & Region III & $ 1.465 $ & $ 0.209 $ & 26 \\
  \hline

  \end{tabular*}
\end{table*}

\begin{table*}[ht!]
\small
  \caption{\textcolor{blue}{Data of crossover times corresponding to Fig.\ref{fig:v-frac-Lxy} for ansitropy $D=+1.0$ in the presence external field $h_{ext}=-0.20$.}}
  \label{tab:crosstime}
  \begin{tabular*}{\textwidth}{@{\extracolsep{\fill}}ccccc}
  \hline
  $L_x = L_y$ & $\ln(t_{c_{1}})$  & 1st Crossover time ($t_{c_{1}}$) & $\ln(t_{c_{2}})$ & 2nd Crossover time ($t_{c_{2}}$)  \\
 \hline
 2 & 2 & Around 8  &  5.306 & Around 201   \\
 4 & 2.25 & Around 9  &  5.754 & Around 316   \\
 8 & 2.296 & Around 10 & 5.299 & Around 200 \\
 16 & 2.13 & Around 9 & 5.02 & Around 152 \\
\hline
 \end{tabular*}
\end{table*}


\begin{table*}[ht!]
\small
  \caption{\textcolor{blue}{Fitting parameters corresponding to Fig.\ref{fig:v-frac-Lxy} for ansitropy $D=+1.0$ in the presence external field $h_{ext}=-0.20$.}}
  \label{tab:cross}
  
  \begin{tabular*}{\textwidth}{@{\extracolsep{\fill}}ccccc}
  \hline
  $L_x = L_y$ & Time regime & $n = a_1\pm \delta a_1$ & $\chi^2$ & DoF \\
 \hline
 
 \multirow{3}{*}{2}
         & Region I & $0.710 \pm 0.030 $  & 0.0187 & 6   \\
         & Region II & $0.189 \pm 0.005 $ & 0.1080 & 108 \\
         & Region III & $1.695 \pm 0.002 $ & 3.5034 & 1962 \\
 \hline
 
 \multirow{3}{*}{4}
         & Region I & $0.723 \pm 0.028 $ & 0.0236 & 7  \\
         & Region II & $0.112 \pm 0.002 $ & 0.0970 & 187  \\
         & Region III & $1.462 \pm 0.002 $ & 4.6980 & 1876  \\
 \hline
 
 \multirow{3}{*}{8}
         & Region I & $0.746 \pm 0.026 $ & 0.0194 & 7  \\
         & Region II & $0.183 \pm 0.003 $ & 0.0672 & 151  \\
         & Region III & $1.642 \pm 0.003 $ & 0.3591 & 512 \\
  \hline 
  
  \multirow{3}{*}{16}
         & Region I & $0.780 \pm 0.024 $ & 0.0083 & 5  \\
         & Region II & $0.222 \pm 0.003 $ & 0.0187 & 88  \\
         & Region III & $2.438 \pm 0.020 $ & 1.2141 & 215  \\
   \hline 
   
  \end{tabular*}
\end{table*}

\begin{table*}[ht!]
\small
  \caption{Fitting parameters corresponding to Fig. \ref{fig:reversaltime}(b) }
  \label{tab:table5}
  \begin{tabular*}{\textwidth}{@{\extracolsep{\fill}}cccc}
  \hline
  External field ($h_{ext}$) & Slope = $b_1 \pm \delta b_1$ & $\chi^2$ & DoF \\
 \hline
  -0.20 & $ -1.208 \pm 0.069 $ & 0.0825 & 4 \\
  -0.25 & $ -1.032 \pm 0.041 $& 0.0295 & 4 \\
\hline
   
  \end{tabular*}
\end{table*}

\end{document}